\documentclass[a4paper,prx,notitlepage,onecolumn,nofootinbib,superscriptaddress,longbibliography,floatfix,11pt,accepted=2026-06-13]{quantumarticle}
\pdfoutput=1
\usepackage[numbers,sort&compress]{natbib}
\usepackage{caption}
\usepackage{subcaption}
\usepackage{amsmath, amssymb, amsfonts, amsthm}
\usepackage{tabularx,graphicx}
\usepackage{epstopdf}
\usepackage{graphicx}
 \usepackage{spectralsequences} 
\usepackage{latexsym}
\usepackage{tikz-cd}
\usepackage{color, colortbl}
\usepackage{psfrag}
\usepackage{bbm}
\usepackage{bm}
\usepackage{dsfont}
\usepackage{feynmp}
\usepackage{slashed}
\usepackage{multirow}
\usepackage{framed}
\usepackage{braket}
\usepackage{color}
\usepackage{hyperref}
\usepackage[normalem]{ulem}

\makeatletter
\def\l@subsubsection#1#2{}
\makeatother

\DeclareMathOperator{\Tr}{Tr}

\theoremstyle{plain}
\newtheorem*{theorem*}{Theorem}

\newcommand{\ident}{\text{\usefont{U}{bbold}{m}{n}1}}
\DeclareMathOperator{\Var}{Var}

\begin{document}
	\title{Measuring R\'enyi entropy with an Echo Protocol}

	\author{Yi-Neng Zhou}
\affiliation{Department of Theoretical Physics, University of Geneva, 24 quai Ernest-Ansermet, 1211 Genève 4, Suisse}
\email{zhouyn.physics@gmail.com}
\orcid{0000-0003-1690-4787}

\author{Robin L\"{o}wenberg}
\email{robin.loewenberg@unige.ch}
\affiliation{Department of Theoretical Physics, University of Geneva, 24 quai Ernest-Ansermet, 1211 Genève 4, Suisse}

\author{Julian Sonner}
\email{julian.sonner@unige.ch}
\affiliation{Department of Theoretical Physics, University of Geneva, 24 quai Ernest-Ansermet, 1211 Genève 4, Suisse}
\affiliation{Jefferson Physical Laboratory, Harvard University, Cambridge, MA 02138, USA}


\begin{abstract}
We present efficient and practical protocols to measure the second R\'enyi entropy, whose exponential is known as the purity. 
Our approach is based on expressing the purity in terms of transition probabilities generated by an echo-type forward-backward evolution sequence, making it applicable to quantum many-body systems. Notably, our approach does not rely on random-noise averaging, a feature that can be extended to protocols to measure out-of-time-order correlation functions, as we demonstrate.  By way of example, we show that our protocols can be practically implemented in superconducting qubit-based platforms, as well as in cavity-QED trapped ultra-cold gases.
    
\end{abstract}

\maketitle

    \tableofcontents


\bigskip

\section{Introduction}
Entanglement entropy, a quantitative measure of entanglement and nonlocal quantum correlations, is a key concept in quantum many-body systems \cite{Horodecki_2009}. In general, it is often characterized as the entropy of the reduced density matrix, which arises in a subsystem when information about the remaining system is ignored and thus traced out. This measure reflects the nonlocal correlations between two parts of the system that are inaccessible through local measurements performed on only one part. The concept of entanglement entropy is broadly significant across various fields, including condensed matter physics \cite{RevModPhys.80.517, Eisert_2010,Laflorencie_2016, Abanin_2019}, quantum information science \cite{Nielsen_Chuang_2010, Plenio:2007zz}, and quantum gravity and high-energy field theory \cite{Pasquale_Calabrese_2004, Ryu_2006, Calabrese_2009, RevModPhys.90.035007}. For example, in condensed matter physics, entanglement entropy serves as a tool to probe quantum criticality \cite{Osborne_2002, Osterloh_2002, Vidal_2003} and non-equilibrium dynamics \cite{Bardarson_2012, Daley_2012}. It also helps to determine the feasibility and efficiency of numerical techniques for studying quantum many-body physics \cite{Schuch_2008}. Furthermore, the notions of entanglement spectrum and entanglement entropy provide a general framework for diagnosing topological phases \cite{Kitaev_2006, Levin_2006, Jiang_2012}. Additionally, entanglement entropy is directly related to other important quantities, such as the out-of-time-order correlator (OTOC) \cite{Fan_2017}, which is a key concept in the study of quantum chaos \cite{Maldacena:2015waa} and quantum gravity \cite{Shenker:2013pqa}. 

Due to its theoretical significance across so many different areas of physics, the experimental measurement of entanglement entropy is evidently of great importance. However, directly measuring entanglement in experiments is extremely challenging. Nevertheless, recent advances in experimental techniques for realizing and controlling quantum simulations have made the measurement of entanglement entropy feasible. In recent years, entanglement entropy has been successfully measured in various platforms, including optical lattices \cite{Islam_2015, Kaufman_2016}, photonic systems \cite{Lin_2024}, trapped-ion platforms \cite{Linke_2018, BrydgesEtAl2019ProbingRenyi}, and ultracold atom simulators \cite{Tajik_2023}. Directly measuring the entanglement entropy of larger systems remains a significant challenge. Existing protocols for measuring entanglement entropy typically require either the preparation of two copies of the system and performing measurements on all sites or the use of randomized measurement techniques. While the latter requires only a single copy of the system, it comes at the cost of implementing the required source of randomness, for example, in the form of a random unitary $ k$-design, which can be highly resource-demanding when the system size is large. Both approaches become increasingly difficult when studying systems of larger sizes. This raises the question: Can we develop a general protocol for measuring entanglement entropy that is both practical and scalable for large systems?

In this paper, we propose a satisfactory answer to this question by developing an experimentally practical protocol for measuring entanglement entropy using an echo-type evolution sequence. Our main result is a direct relation between the purity, namely the exponential of the second R\'enyi entropy, and a sum of basis-resolved echo transition probability (ETP) generated by forward and backward time evolution. This relation allows us to convert the measurement of entanglement entropy into the measurement of transition probabilities in an echo-type protocol, which is directly applicable to quantum many-body systems. The protocol can be implemented on experimental platforms that support echo-type measurements, including superconducting qubits \cite{Braum_ller_2021}, NMR systems \cite{Li_2017}, and cavity QED (cQED) systems capable of generating Hamiltonians with holographic duals. This makes the protocol especially interesting in the context of holographic duality, where the evolution of entanglement entropy during black hole evaporation plays a central role in the understanding of the Page curve and the unitarization of Hawking radiation \cite{Penington:2019npb,Almheiri:2019psf,Almheiri:2020cfm}. From the replica-saddle-point perspective, the purity itself is expected to display Page-curve behavior \cite{Almheiri:2019qdq}, making it a natural and experimentally relevant observable.

The outline of the paper is as follows: in Section \ref{RE-LE_relation}, we introduce the relation between the second R\'enyi (entanglement) entropy and the ETP. In Section \ref{LE_experiment}, we define the basis-resolved ETP and propose the experimental protocol for its measurement, showing that the sum of the basis-resolved ETP provides the quantum purity, which is directly related to the second R\'enyi entropy. In Section \ref{OTOC-LE_relation_main}, we give a diagrammatic proof of the OTOC-LE relation, bypassing the need for a random noise ensemble average. In Section \ref{application}, we present two experimental applications of our R\'enyi-entropy measurement protocol. The first is in superconducting circuits and is illustrated by a three-qubit example; the second is in cavity QED implementations, including a recent proposal for simulating p-adic AdS/CFT.
 In Section \ref{summary}, we provide a summary of our paper and discuss some advantages of our experimental proposal over previous ones, as well as its theoretical implications.

\section{A Relation between R\'enyi Entropy and Echo Transition Probability}
\label{RE-LE_relation}
In this section, we examine the relation between R\'enyi entropy and echo transition probabilities, focusing first on the second R\'enyi entropy for simplicity. Based on this relation, we then propose an experimental protocol for measuring R\'enyi entropies using an echo-type evolution sequence.

We begin by introducing the second R\'enyi entropy and the Loschmidt echo. The second R\'enyi entropy is defined in terms of the purity as
\begin{equation}
S^{(2)}=-\log\!\left[\Tr(\hat{\rho}^{\,2})\right].
\end{equation}
The Loschmidt echo (LE) \cite{PhysRevA.30.1610,PASTAWSKI2000166,PhysRevLett.86.2490,gorin2006dynamics,doi:10.1080/00018730902831009,goussev2012loschmidt} is defined as
\begin{equation}
M(t)=\left|\langle \psi_0|e^{i\hat{H}_2 t}e^{-i\hat{H}_1 t}|\psi_0\rangle\right|^2.
\label{Echo_closed_def}
\end{equation}
Here, $\hat{H}_1$ and $\hat{H}_2$ govern the forward and backward time evolution, respectively, and $|\psi_0\rangle$ is the initial state at time $t_0=0$. When $\hat{H}_2=\hat{H}_1+\hat{V}$, with $\hat{V}$ a perturbation to $\hat{H}_1$, the Loschmidt echo quantifies the sensitivity of the dynamics to the perturbation and thus serves as a probe of irreversibility.

More generally, one may consider an echo-type protocol in which the final state onto which one projects is not necessarily the initial state. This motivates the notion of an echo transition probability (ETP),
\begin{equation}
P_{\phi \leftarrow \psi_0}(t)
=
\left|
\langle \phi|e^{i\hat{H}_2 t}e^{-i\hat{H}_1 t}|\psi_0\rangle
\right|^2,
\label{general_echo_transition}
\end{equation}
which gives the probability that the forward-backward evolution maps the initial state $|\psi_0\rangle$ to a target state $|\phi\rangle$. The Loschmidt echo in Eq.~\eqref{Echo_closed_def} is recovered as the special case in which the target state coincides with the initial state, \emph{i.e.}, $|\phi\rangle=|\psi_0\rangle$. In this sense, the Loschmidt echo is a particular example of an echo transition probability.

\begin{figure}[t]
		\centering \includegraphics[width=0.9\textwidth]{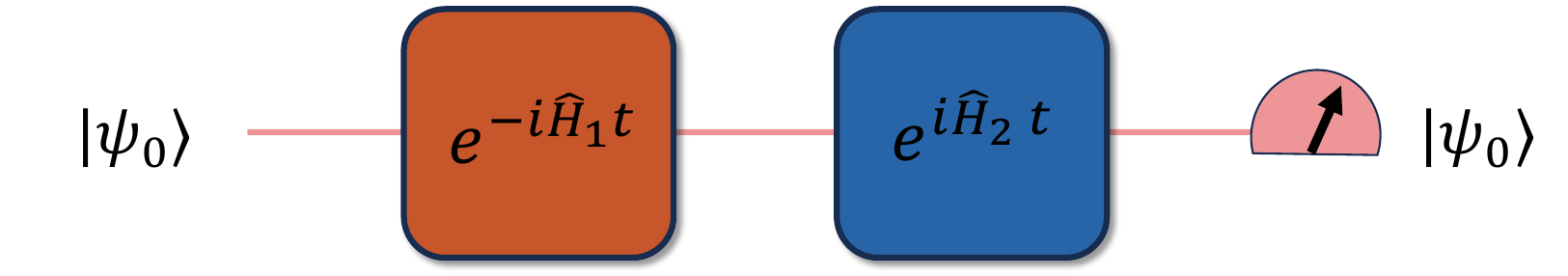}
		\caption{Protocol for measuring the Loschmidt echo defined in Eq.~\eqref{Echo_closed_def}. The protocol consists of forward evolution followed by backward evolution and a final projection onto the initial state. This return-probability measurement is a special case of the more general echo transition probabilities introduced in Eq.~\eqref{general_echo_transition}. Note that the protocol involves a time-inversion step, which we discuss further below.}
		\label{LE_protocol}
	\end{figure}

The protocol for measuring the Loschmidt echo defined in Eq.~\eqref{Echo_closed_def} is shown in Fig.~\ref{LE_protocol}. As we will show below, the quantities entering our R\'enyi entropy protocol are not Loschmidt echo in the strict sense, but rather basis-resolved echo transition probabilities of the form in Eq.~\eqref{general_echo_transition}. This leads to the central relation in this work: a relation between the second R\'enyi entropy and a sum of echo transition probabilities.

Below, we develop a relation between the second R\'enyi entropy and the ETP. Consider a scenario where the total system is partitioned into subsystems A and B. The time evolution of the reduced density matrix for subsystem A is given by
\begin{equation}
    \label{reduced_evolution}
    \hat{\rho}_A(t) = \text{Tr}_B \left[ \hat{U}(t) \hat{\rho}(0) \hat{U}^{\dagger}(t) \right].
\end{equation}
Here, $\hat{U}(t)=e^{-i\hat{H}t}$ is the unitary time evolution of the total system.

Assuming the initial state is the product state of subsystems A and B and that the initial states for subsystems A and B are both pure, we can denote the initial density operator as
\begin{equation}
\hat{\rho}(0) = |\psi_0\rangle_{AA} \langle \psi_0| \otimes |B_0\rangle_{BB} \langle B_0|.
\end{equation}
Here, $|\psi_0\rangle $ and $|B_0\rangle$ are the (pure) initial states for subsystems A and B, respectively.

The purity can be directly rewritten using the definition in Eq.~\eqref{reduced_evolution} as  
\begin{equation} 
\label{purity_projected_LE}
\begin{aligned}
    &\text{Tr}_A \left[\hat{\rho}_A^2(t) \right] \\
    =&\text{Tr}_A \left\{ \text{Tr}_{B_1}\left[\hat{U}_{A,B_1}(t)\hat{\rho}_A^0 \otimes \hat{\rho}_{B_1}^0\hat{U}^{\dagger}_{A,B_1}(t)\right] \text{Tr}_{B_2}\left[ \hat{U}_{A,B_2}(t)\hat{\rho}_A^0 \otimes \hat{\rho}_{B_2}^0\hat{U}^{\dagger}_{A,B_2}(t) \right]\right\}\\
    =&\text{Tr}_{A\cup B_1 \cup B_2} \left[  \hat{\rho}_A^0 \otimes \hat{\rho}_{B_1}^0 \otimes \hat{\ident}_{B_2} \hat{U}^{\dagger}_{A,B_1}   \hat{U}_{A,B_2} \hat{\rho}_A^0 \otimes \hat{\ident}_{B_1} \otimes \hat{\rho}_{B_2}^0 \hat{U}^{\dagger}_{A,B_2}  \hat{U}_{A,B_1}  \right]\\
=& \sum_{m_1,m_2=1}^{D_B} \left| \langle \psi_0, B_0, m_2 | \hat{U}_{A,B_1}^\dagger  \hat{U}_{A,B_2}  | \psi_0, m_1, B_0 \rangle \right|^2.
\end{aligned}
\end{equation}
For simplicity of notation, we omit the temporal argument in the unitary evolution operators from the third line onward. Here, $\hat{\ident}_B$ represents the identity operator on subsystem B. From the second to the third line, we have used the cyclic property of the trace. In the fourth line, we use the spectral decompositions $\hat{\ident}_{B_1}=\sum_{m_1=1}^{D_B}|m_1\rangle \langle m_1|$, and $\hat{\ident}_{B_2}=\sum_{m_2=1}^{D_B}|m_2\rangle \langle m_2|$. Here, $m_1 (m_2)$ ($m_1, m_2 = 1, 2, \dots, D_B$) labels the complete orthogonal basis of subsystems $B_1 (B_2)$, and $D_B$ represents the dimension of the Hilbert space of subsystem $B$ (both $B_1$ and $B_2$ have the same Hilbert space dimension). Additionally, $\hat{U}_{A, B_1} (\hat{U}_{A, B_2})$ denotes the unitary evolution of subsystems $A$ and $B_1 (B_2)$ together. Finally, $|\psi_0, B_0, m_2\rangle$ represents $|\psi_0\rangle_A\otimes |B_0\rangle_{B_1} \otimes |m_2\rangle_{B_2}$. 

In the above derivation, the key idea is to introduce two copies of subsystem B ($B_1$ and $B_2$), which differentiates the forward and backward time evolution. Without this distinction, the forward and backward evolutions would cancel each other if both involved the same subsystem B. Since $B_1$ and $B_2$ are independent, this approach allows us to ultimately express the purity in terms of the ETP.  

From this derivation, we find that the purity can be expressed as a sum of transition probabilities associated with an echo-type evolution protocol. The protocol starts from the state $|\psi_0,m_1,B_0\rangle$. The subsystems $A$ and $B_2$ first evolve forward in time for a duration $t$, followed by backward evolution of $A$ and $B_1$ for the same duration. We then project onto the target state $|\psi_0,B_0,m_2\rangle$. The corresponding projection probability is
\begin{equation}
    M(t,m_1,m_2)\equiv
    \left|
    \langle \psi_0,B_0,m_2|
    \hat U_{A,B_1}^{\dagger}(t)\hat U_{A,B_2}(t)
    |\psi_0,m_1,B_0\rangle
    \right|^2.
\label{LE_projeted_def}
\end{equation}
This quantity is the transition probability from the initial state $|\psi_0,m_1,B_0\rangle$ to the target state $|\psi_0,B_0,m_2\rangle$ under a forward-backward (echo-type) evolution sequence. We then called it basis-resolved echo transition probability (ETP). Unlike the conventional Loschmidt echo, which measures the return probability to the same initial state, Eq.~\eqref{LE_projeted_def} describes a basis-resolved transition probability between generally distinct initial and final states. The purity is obtained by summing $M(t,m_1,m_2)$ over $m_1$ and $m_2$. By combining the derivation in Eq.~\eqref{purity_projected_LE} with the definition of the basis-resolved ETP in Eq.~\eqref{LE_projeted_def}, we find that the purity can be expressed as the sum of the basis-resolved ETP
\begin{equation}
    \text{Tr}_A \bigl[\hat{\rho}_A^2(t)\bigr] 
= \sum_{m_1,m_2=1}^{D_B} M(t, m_1, m_2).
\end{equation}
Thus, the second R\'enyi entropy, whose negative is the logarithm of the quantum purity, can be written as
\begin{equation}
    \label{RE-LE_relation_equation}
    S^{(2)}=-\log\bigl[\sum_{m_1,m_2=1}^{D_B} M(t, m_1, m_2)\bigr].
\end{equation}
In the next section, we will propose an experimental protocol for measuring quantum purity, the logarithm of which yields the second R\'enyi entropy. This protocol is similar to the one used for measuring the basis-resolved ETP.

\section{Efficient Protocol for Measuring R\'enyi Entropy}  
\label{LE_experiment}  
In this section, we first introduce a protocol for experimentally measuring the basis-resolved ETP defined in Eq.~\eqref{LE_projeted_def} and then show its direct application to measuring the second R\'enyi entropy.

\subsection{Measurement Protocol for the Echo Transition Probability}
\label{measure_projected_LE_subsection}
 \begin{figure}[t] 
		\centering \includegraphics[width=0.95\textwidth]{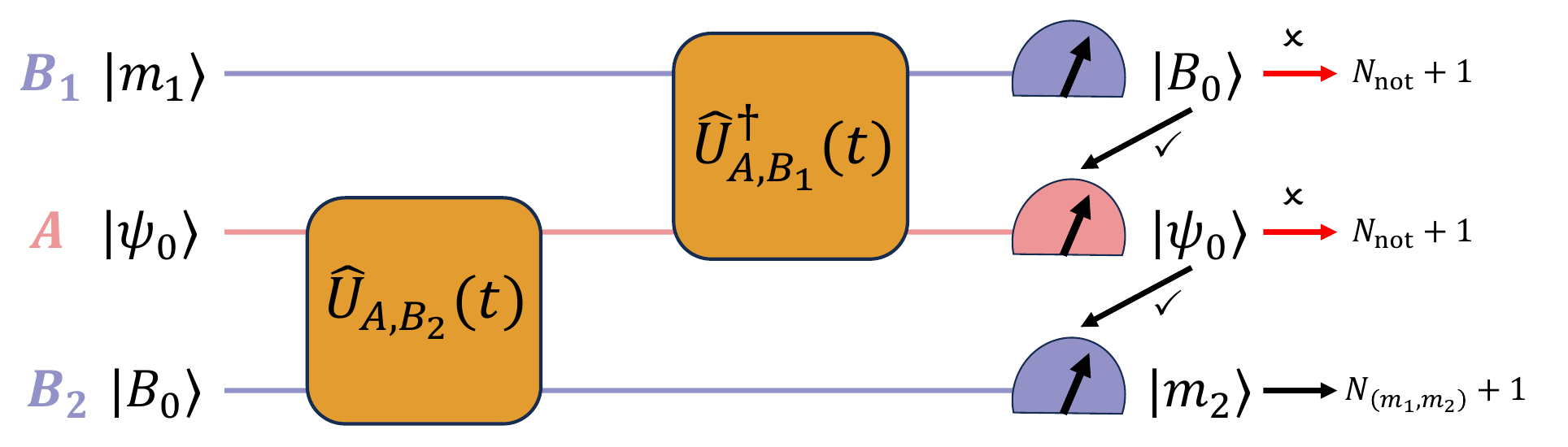} 
		\caption{The protocol for each round of the measurement of basis-resolved ETP $M(t,m_1,m_2)$ defined in Eq.~\eqref{LE_projeted_def}. We start with the initial state $|m_1,\psi_0,B_0\rangle$ and let subsystems $ A $ and $ B_2 $ evolve forward in time for $ t $. Then, we evolve subsystems $ A $ and $B_1$ backward in time for the same duration $ t $. Finally, we perform a measurement on subsystems $ B_1 $,$A$, and $B_2$. The protocol is discussed in detail in subsection \ref{measure_projected_LE_subsection}.}
		\label{projected_LE_protocol}
\end{figure}

We begin by proposing a protocol for measuring the basis-resolved ETP defined in Eq.~\eqref{LE_projeted_def}. As discussed, expressing quantum purity as the basis-resolved ETP requires two copies of subsystem $B$. For practical implementation, we assume $A$ is larger than $B$ and consider a qubit-based system where $A(B)$ consists of $N_A(N_B)$ qubits. Here, for simplicity, we choose $N_B=1$ as an example (although $N_B$ can generally be much larger than $1$). For clarity, we set the initial states as $ |\psi_0\rangle_A = |+1, +1, \dots, +1\rangle_A $ for subsystem $ A $ and $ |B_0\rangle_B = |+1\rangle_B $ for subsystem $ B $ in the $ \hat{\sigma}_z $ basis.

Initially, we set $N_{(|m_1\rangle, |m_2\rangle)} = 0$ for all $m_1, m_2$, where $m_1 (m_2)$ ($m_1,m_2 = 1$, $2$, $\dots$, \\$D_B=2^{N_B}$) denotes the label of the $\hat{\sigma}_z$ measurement outcomes for subsystems $B_1$ and $B_2$, respectively. Here, $|-1\rangle$ corresponds to $|m = 1\rangle$ and $|+1\rangle$ corresponds to $|m = 2\rangle$. 
We initialize $N_{\text{count}} = 0$, $N_{\text{not}} = 0$, and $m_1 = 1$. The proposed protocol for measuring the basis-resolved ETP $M(t, m_1, m_2)$, as defined in Eq.~\eqref{LE_projeted_def}, consists of the following steps:  

\begin{enumerate}
    
    \item  Prepare the initial state of subsystem $A$ as $|+1,+1,\ldots,+1\rangle_A$, and the initial state of subsystems $B_1$ as $|m=m_1\rangle_{B_1}$ and $B_2$ as $|B_0\rangle=|+1\rangle_{B_2}$.
    
    \item  Let subsystem $A$ and $B_2$ evolve unitarily together for time $t$, then let subsystem $A$ and $B_1$ evolve together backward for time $t$.

    \item  Measure $\hat{\sigma}_z$ on the subsystem $B_1$. If the result is not $|+1\rangle_{B_1}$, update $N_{\text{not}} \to N_{\text{not}}+1$, skip steps 4 and 5, and proceed directly to step 6. 

    \item  Measure $\hat{\sigma}_z$ on each qubit of subsystem $A$. If the result is not $|+1,+1,\ldots,+1\rangle_A$, update $N_{\text{not}} \to N_{\text{not}}+1$, skip step 5 and proceed directly to step 6.  

    \item  Measure $\hat{\sigma}_z$ on the subsystem $B_2$ and find the according label of the measurement result $m_2$. Then update $N_{(|m_1\rangle, |m_2\rangle)} \to N_{(|m_1\rangle, |m_2\rangle)} + 1$.

    \item  Update the $N_{\text{count}} \to N_{\text{count}}+1$. If $N_{\text{count}} <N_{\text{cycle}}$, go back to step 1. Otherwise, update $m_1 \to m_1+1$,  $N_{\text{count}} \to 0$, and go back to step 1 if $m_1 < D_B$.

\end{enumerate}
This measurement protocol is shown in Fig.~\ref{projected_LE_protocol}. Here, $N_{\text{cycle}}$ denotes the total number of the rounds of the experiment for each given $m_1$, $N_{\text{not}}$ denote the total number of rounds during which subsystem A and $B_1$ do not return to the given final state $|\psi_0\rangle$ and $|B_0\rangle$,  and $ N_{(|m_1\rangle, |m_2\rangle)} $ denotes the total number of rounds that start from the initial state $ |\psi_0, m_1, B_0\rangle $ and, after forward and backward time evolution, result in the final state $ |\psi_0, B_0, m_2\rangle $. From the above steps, one can compute the basis-resolved ETP for each pair of labels $(m_1,m_2)$:
\begin{equation}
 M(t,m_1,m_2)=\frac{N_{(|m_1\rangle, |m_2\rangle)}}{N_{\text{cycle}}}.
\end{equation}
The use of having kept track of $N_{\rm not}$ will become clear when we measure R\'enyi entropy.

\subsection{Measuring the Second R\'enyi Entropy}
\label{measure_Renyi_entropy_subsection}
We now discuss how to measure the second R\'enyi entropy, building on the protocol we proposed for measuring the basis-resolved ETP. We first present the general proposal, which can be directly inferred from our discussion of the measurement of the basis-resolved ETP in the previous section. Subsequently, we generalize our protocol to a special case where the experiment always starts from a fixed state of system B, which may be more practical when the size of subsystem B is very large.

\subsubsection{General Proposal: Averaging Over All Possible States of Subsystem B}
\label{sec:general_proposal}
 As discussed in the previous section, the purity can be expressed as the sum of the basis-resolved ETP,
\begin{equation}
\label{purity_formula}
\begin{aligned}
    \text{Tr}_A [\hat{\rho}_A^2(t)]
    =\sum_{m_1,m_2=1}^{D_B}M(t,m_1,m_2)=\frac{1}{N_{\text{cycle}}}\sum_{m_1,m_2=1}^{D_B} N_{(|m_1\rangle, |m_2\rangle)}.
\end{aligned}
\end{equation}
From the protocol for measuring basis-resolved ETP discussed in the previous subsection, we have
\begin{equation}
    N_{\text{not}}=D_BN_{\text{cycle}}-\sum_{m_1,m_2=1}^{D_B}N_{(|m_1\rangle, |m_2\rangle)}.
\end{equation}
Thus, by combining the above equation with Eq.~\eqref{purity_formula}, the purity can be further rewritten as 
\begin{equation}
    \text{Tr}_A [\hat{\rho}_A^2(t)]
    =D_B-\frac{N_{\text{not}}}{N_{\text{cycle}}}.
\end{equation}
The second R\'enyi entropy is
\begin{equation}
S^{(2)}=-\log\left[D_B-\frac{N_{\text{not}}}{N_{\text{cycle}}}\right].
\label{Renyi_entropy_protocol_equation}
\end{equation}   
From the above formula, we can see that if one is interested in measuring the second R\'enyi entropy, it is sufficient to count the number of $N_{\text{not}}$ in the protocol, without needing to know the exact value of $N_{(|m_1\rangle, |m_2\rangle)}$ for each pair of $(m_1, m_2)$. The measurement protocol for the second R\'enyi entropy can thus be further simplified compared to the previous protocol by starting with $N_{\text{not}} = 0$, $N_{\text{count}} = 0$, and following the steps outlined below:  

\begin{figure}[t] 
    \centering \includegraphics[width=0.95\textwidth]{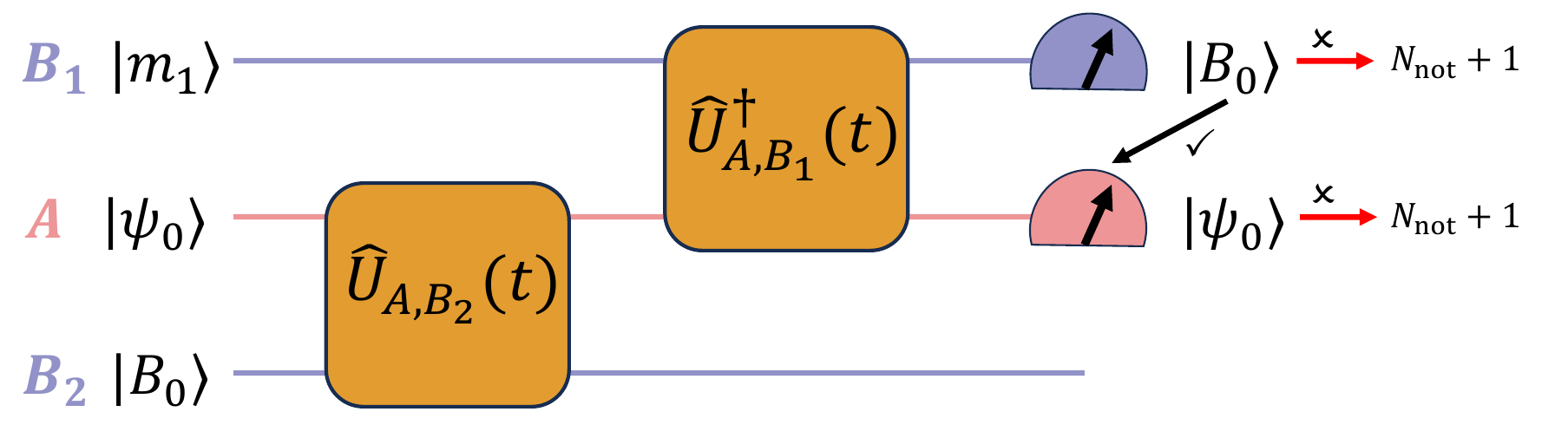} 
    \caption{Single-round protocol for measuring the second R\'enyi entropy via the basis-resolved ETP begins with the given initial state $|m_1,\psi_0,B_0\rangle$. We start with the initial state $|m_1,\psi_0,B_0\rangle$ and let subsystems $ A $ and $ B_2 $ evolve forward in time for $ t $. Then, we evolve subsystems $ A $ and $ B_1 $ backward in time for the same duration $ t $. Finally, we perform a measurement on subsystems $ B_1 $ and $ A $. This protocol is discussed in detail in subsection \ref{sec:general_proposal}.}
    \label{Renyi2_protocol}
\end{figure}
\begin{enumerate}
    
    \item  Prepare the initial state of subsystem $A$ as $|+1,+1,\ldots,+1\rangle_A$, and the initial state of subsystems $B_1$ as $|m=m_1\rangle_{B_1}$ and $B_2$ as $|B_0\rangle=|+1\rangle_{B_2}$.
    
    \item  Let subsystem $A$ and $B_2$ to evolve unitarily together for time $t$, then let subsystem $A$ and $B_1$ evolve together backward for time $t$.
    
    \item  Measure $\hat{\sigma}_z$ on subsystems $B_1$. If the result is not $|+1\rangle_{B_1}$, update $N_{\text{not}} \to N_{\text{not}}+1$, skip steps 4 and proceed directly to step 5. 
    
    \item  Measure $\hat{\sigma}_z$ on each qubit of subsystem $A$. If the result is not $|+1,+1,\ldots,+1\rangle_A$, update $N_{\text{not}} \to N_{\text{not}}+1$. 
    
    \item  Update the $N_{\text{count}} \to N_{\text{count}}+1$. If $N_{\text{count}} <N_{\text{cycle}}$, go back to step 1. Otherwise, update $m_1 \to m_1+1$, and $N_{\text{count}} \to 0$. Go back to step 1 if $m_1 < D_{B}$.

\end{enumerate}
The measurement protocol is visualized in Fig.~\ref{Renyi2_protocol}.
The protocol described above for measuring the second R\'enyi entropy requires approximately $\sim 2^{N_B} N_B N_A 
\times N_{\rm cycle}$ individual measurements. In comparison, the protocol used to measure the specific basis-resolved ETP requires approximately $\sim  N_B^2 N_A 
\times N_{\rm cycle}$ individual measurements.

\subsubsection{Simplified Version: Need Only One Copy of A and B.}
\label{sec:simplified_fixed_approach}
\begin{figure}[t] 
    \centering \includegraphics[width=0.95\textwidth]{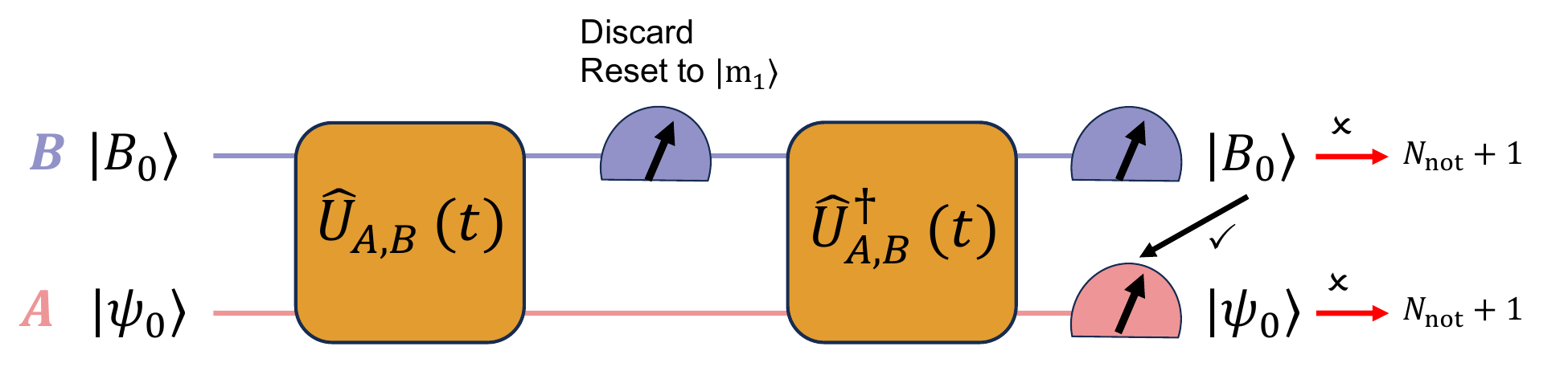} 
    \caption{Simplified single-round protocol for measuring the second R\'enyi entropy via the basis-resolved ETP with a \emph{given} initial state. Starting from $\ket{\psi_0}_A\otimes\ket{B_0}_B$, we evolve $A$ and $B$ forward for time $t$, discard the final state of $B$, reset $B$ to $\ket{m_1}$, and then evolve $A$ and $B$ backward for the same duration $t$. Finally, we measure subsystems $A$ and $B$. See Sec.~\ref{sec:simplified_fixed_approach} for details.}
    \label{Renyi2_protocol_simplified}
\end{figure}

Drawing on ideas from dissipative state preparation \cite{ ding2025endtoendefficientquantumthermal} where an ancilla can be reused to monitor open system dynamics, we can further simplify the measurement protocol for the basis-resolved ETP. Specifically, we treat subsystem $B$ as the ancilla and subsystem $A$ as the open system. After the forward time evolution, we reset $B$ to the required initial state and reuse it for the backward evolution. This removes the need for two copies of $B$: a single ancilla suffices to measure the basis-resolved ETP. The simplified protocol is shown in Fig.~\ref{Renyi2_protocol_simplified}.

The measurement protocol for the second R\'enyi entropy can thus be further simplified compared to the previous protocol by starting with $N_{\text{not}} = 0$, $N_{\text{count}} = 0$, and following the steps outlined below: 

\begin{enumerate}
    
    \item  Prepare the initial state of subsystem $A$ as $|+1,+1,\ldots,+1\rangle_A$, and the initial state of subsystems $B$ as $|B_0\rangle=|+1\rangle_{B}$.
    
    \item  Let subsystem $A$ and $B$ to evolve unitarily together for time $t$, 
    
    \item Reset the state of $B$ to be $|m=m_1\rangle_B$, then let subsystem $A$ and $B$ evolve together backward for time $t$.
    
    \item  Measure $\hat{\sigma}_z$ on subsystems $B$. If the result is not $|+1\rangle_{B}$, update $N_{\text{not}} \to N_{\text{not}}+1$, skip steps 5 and proceed directly to step 6. 
    
    \item  Measure $\hat{\sigma}_z$ on each qubit of subsystem $A$. If the result is not $|+1,+1,\ldots,+1\rangle_A$, update $N_{\text{not}} \to N_{\text{not}}+1$. 
    
    \item  Update the $N_{\text{count}} \to N_{\text{count}}+1$. If $N_{\text{count}} <N_{\text{cycle}}$, go back to step 1. Otherwise, update $m_1 \to m_1+1$, and $N_{\text{count}} \to 0$. Go back to step 1 if $m_1 < D_{B}$.

\end{enumerate}

The second R\'enyi entropy is also given by Eq.~\eqref{Renyi_entropy_protocol_equation}.

\subsubsection{Random Unitary Approach: Initializing from a Fixed State of Subsystem B}
\label{sec:random_unitary_approach}
In the above protocol, all states from a complete orthogonal basis of subsystem $B$ are required to be prepared (one by one) as initial states for the evolution, as their basis-resolved ETP sum to the desired second R\'enyi entropy. However, this can be challenging when the size of the subsystem $B$ is large. Here, we assume that $N_B$, the number of qubits in subsystem B, is much larger than $1$. This difficulty can be addressed by preparing a fixed initial state for $B_1$ in step 1 and then applying a random unitary rotation to subsystem $B_1$ before step 2.  

The measurement protocol for the second R\'enyi entropy can thus be modified by starting with $\widetilde{N}_{\text{not}} = 0$, $N_{\text{count}} = 0$, and following the steps outlined below:

\begin{enumerate}
    
    \item  Prepare the initial state of subsystem $A$ as $|+1,+1,\ldots,+1\rangle_A$, and the initial state of subsystems $B$ as $|+1,+1,\ldots,+1\rangle_B$.
    
    \item  Let subsystem $A$ and $B$ to evolve unitarily together for time $t$, 

    \item  Apply a random unitary rotation $\hat{u}_B$ on subsystem $B$.
    
    \item Let subsystem $A$ and $B$ evolve together backward for time $t$.

    \item  Measure $\hat{\sigma}_z$ on all the qubits in subsystem $B$. If the result is not $|+1,+1,\ldots,+1\rangle_{B}$, update $\widetilde{N}_{\text{not}} \to \widetilde{N}_{\text{not}}+1$, skip steps 5 and proceed directly to step 6. 
    
    \item  Measure $\hat{\sigma}_z$ on each qubit of subsystem $A$. If the result is not $|+1,+1,\ldots,+1\rangle_A$, update $\widetilde{N}_{\text{not}} \to \widetilde{N}_{\text{not}}+1$. 
    
    \item  Update the $N_{\text{count}} \to N_{\text{count}}+1$. If $N_{\text{count}} <N_{\text{total}}$, go back to step 1.
\end{enumerate}

\begin{figure}[t] 
    \centering \includegraphics[width=0.95\textwidth]{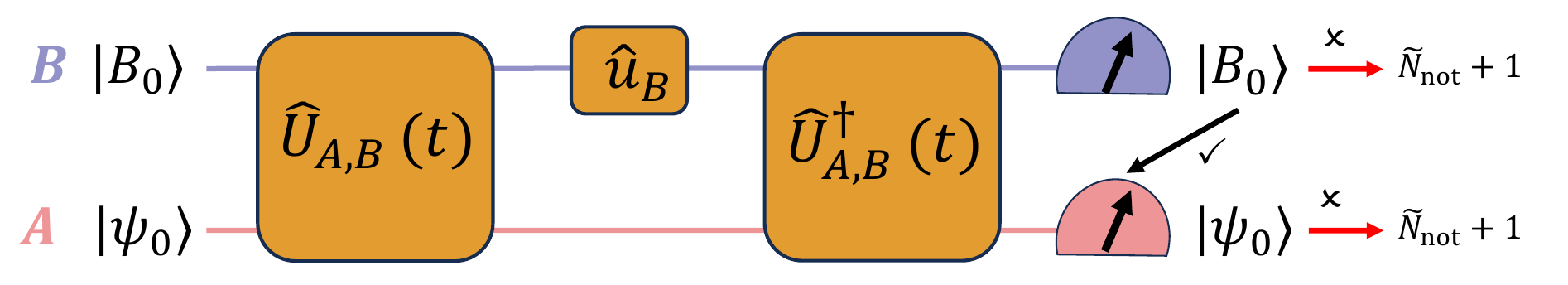} 
    \caption{Single-round protocol for measuring the second R\'enyi entropy via the basis-resolved ETP with a \emph{fixed} initial state each round. Starting from $\ket{\psi_0}_A\otimes\ket{B_0}_B$, we evolve $A$ and $B$ forward for time $t$, apply a random unitary on subsystem $B$, and then evolve $A$ and $B$ backward for the same duration $t$. Finally, we measure subsystems $A$ and $B$. See Sec.~\ref{sec:random_unitary_approach} for details.}
\label{Renyi2_protocol_random_simplified}
\end{figure}

The measurement protocol is shown in Fig.~\ref{Renyi2_protocol_random_simplified}. Here, $N_{\text{total}}$ denotes the total number of rounds of the experiment. We now add a few more details underlying the above protocol.

Since the purity can be expressed as the sum of the basis-resolved ETP, in the protocol above, it can be written as
\begin{equation}
\label{purity_formula_random}
    \text{Tr}_A [\hat{\rho}_A^2(t)]
    =\frac{D_B}{N_{\text{total}}}\int d\hat{u}_1\sum_{m_2=1}^{D_B} N_{(\hat{u}_1|B_0\rangle,|m_2\rangle)}.
\end{equation}
It is obtained from substituting $\sum_{m_1} \to D_B\int d\hat{u}_1 $ in Eq.~\eqref{purity_formula}. Here, the distribution of the random unitary $\hat{u}_1$ satisfies the definitions of a unitary $1-$design. From the above protocol, we have
\begin{equation}
    \widetilde{N}_{\text{not}}=N_{\text{total}}-\int d\hat{u}_1\sum_{m_2=1}^{D_B}N_{(\hat{u}_1|B_0\rangle,|m_2\rangle)}.
\end{equation}
Thus, by combining the above equation with Eq.~\eqref{purity_formula_random}, the purity can be further rewritten as
\begin{equation}
    \text{Tr}_A [\hat{\rho}_A^2(t)]=D_B(1-\frac{\widetilde{N}_{\text{not}}}{N_{\text{total}}}).
\end{equation}
Then, the second R\'enyi entropy can be computed as
\begin{equation}
S^{(2)}=-\log[D_B(1-\frac{\widetilde{N}_{\text{not}}}{N_{\text{total}}})].
\end{equation}

\textit{Remarks on time reversal}: The measurement protocol for the basis-resolved ETP here requires time reversal. However, this requirement can be circumvented by using an alternative approach based on randomized measurements, similar to what was done for OTO-correlations in \cite{nie2019detectingscramblingstatisticalcorrelations, Joshi_2020}. We discuss the time-reversal-free R\'enyi entropy measurement protocol in Appendix \ref{measure_RE_without_time_reversal}. \footnote{ \textit{Remarks on Imperfect time reversal:} Imperfect time reversal is the main experimental bottleneck of echo-based protocols. In practice, one can benchmark the achievable time window by measuring the return probability $\mathcal{L}(t)=\big|\langle\psi_0|\hat{E}(t)|\psi_0\rangle\big|^2$ with $\hat{E}(t)=\hat{\mathcal{T}}\exp\!\Big(i\int_0^t ds\,\delta \hat{H}_I(s)\Big)$ and then restrict to times for which $\mathcal{L}(t)$ remains close to unity, so that Eq.~\eqref{eq:signal_bound_by_echo} ensures controlled systematics. The observed contrast loss can be folded into conservative error bars. Further details are given in Appendix~E. }

\section{OTOC--ETP Relation without Random Noise Average}
\label{OTOC-LE_relation_main}
The R\'enyi-entropy-ETP relation from Section \ref{RE-LE_relation} naturally extends to an OTOC-ETP relation without requiring a random noise average, unlike \cite{yanInformationScramblingLoschmidt2020}\footnote{The random noise approximation in \cite{yanInformationScramblingLoschmidt2020} accounts for the coupling's effect on the reduced evolution of $\hat{W}_A$ by treating it as random noise acting on A:
\begin{equation*}
	\text{Tr}_A(e^{i\hat{H}t}\hat{W}_Ae^{-i\hat{H}t}) \simeq D_A\overline{e^{i(\hat{H}_A +\hat{V}_{\alpha})t}\hat{W}_Ae^{-i(\hat{H}_A +\hat{V}_{\alpha})t}}.
 \label{reduced_approx}
\end{equation*} 
Here, $\{\hat{V}_{\alpha}\}$ represents the random noise operator, with the overline denoting an average over its realizations. This approximation follows from viewing the reduced evolution of $\hat{W}_A$ as an open system dynamics. Under the Born-Markov approximation, it follows the Lindblad master equation, where the jump operators are determined by system-bath interactions. Equivalently, this evolution can be described as system dynamics under an effective Hamiltonian with random noise. The ensemble of this noise, known as Langevin noise, is constrained by the interaction between subsystems A and B.
 }. Using a diagrammatic approach similar to \cite{fanOutofTimeOrderCorrelationManyBody2017, Mele_2024}, we derive this relation without relying on a random noise ensemble for $\hat{\rho}_A$. Instead, it only uses the Haar random average over the subsystem being traced out to represent the time evolution of the other operator, which is not randomly averaged, as the reduced density matrix.

We first briefly review the diagrammatic derivation of the OTOC--R\'enyi-entropy relation~\cite{fanOutofTimeOrderCorrelationManyBody2017}. Combining this with the R\'enyi-entropy--ETP relation established in Sec.~\ref{RE-LE_relation}, we then obtain an OTOC-ETP relation that does \emph{not} require noise-ensemble averaging.

\begin{figure}[t]
    \centering
    \includegraphics[width=0.95\textwidth]{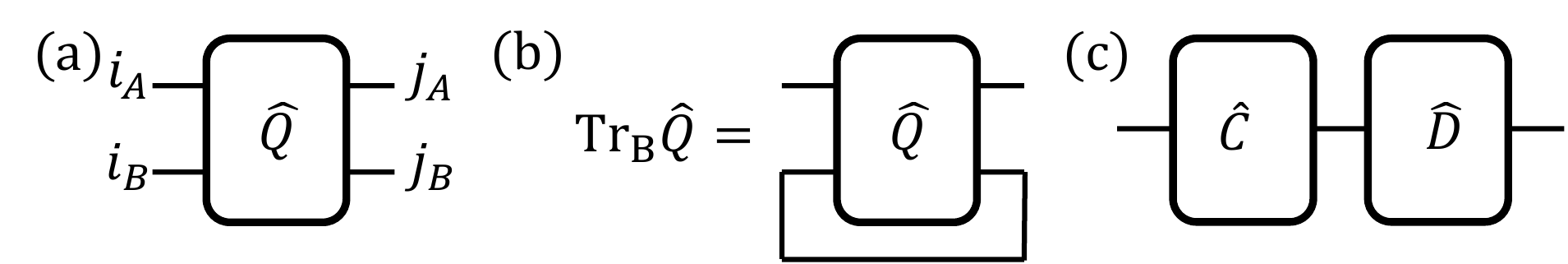}
    \caption{Diagrammatic notation used in Sec.~4.
    (a) Operator $\hat Q$ as a box with input legs $(i_A,i_B)$ and output legs $(j_A,j_B)$.
    (b) Partial trace over $B$ corresponds to contracting the $B$ legs.
    (c) Operator product $\hat C\hat D$ is represented by concatenation and contraction of matching legs.}
    \label{operator_trace_diagram}
\end{figure}

\paragraph{Diagrammatic notation.}
We partition the total system into subsystems $A$ and $B$. For an operator $\hat Q$ acting on $\mathcal{H}_A\otimes\mathcal{H}_B$, and orthonormal bases $\{|i_A\rangle\}$ and $\{|i_B\rangle\}$, we write
\begin{equation}
\hat Q=\sum_{i_A,i_B,j_A,j_B} Q_{i_A,i_B;\,j_A,j_B}\,
|i_A\rangle|i_B\rangle\langle j_A|\langle j_B|.
\end{equation}
Graphically, $\hat Q$ is a box with \emph{input} legs $(i_A,i_B)$ on the left and \emph{output} legs $(j_A,j_B)$ on the right [Fig.~\ref{operator_trace_diagram}(a)].
A partial trace over $B$ is represented by contracting the $B$ legs,
\begin{equation}
\mathrm{Tr}_B\,\hat Q
=\sum_{i_B}\langle i_B|\hat Q|i_B\rangle,
\end{equation}
as in Fig.~\ref{operator_trace_diagram}(b).
The product $\hat C\hat D$ is obtained by placing $\hat C$ to the left of $\hat D$ and contracting the output legs of $\hat C$ with the corresponding input legs of $\hat D$ [Fig.~\ref{operator_trace_diagram}(c)].

\begin{figure}[t]
    \centering
    \includegraphics[width=0.85\textwidth]{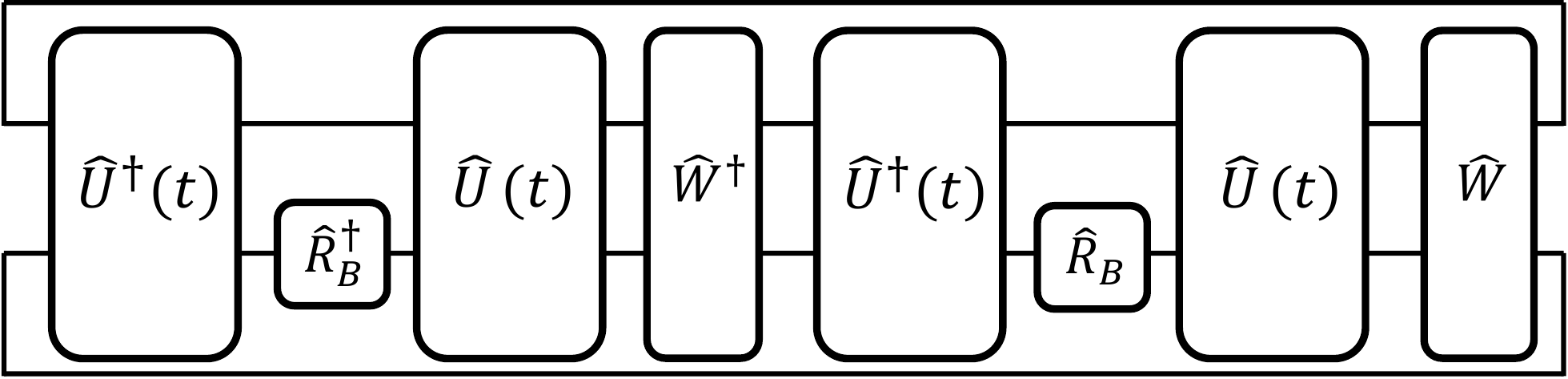}
    \caption{Diagrammatic representation of the OTOC in Eq.~\eqref{OTOC_infinite_def}. Operators are boxes with input (left) and output (right) legs; upper (lower) legs correspond to subsystem $A$ ($B$). Operator multiplication is depicted by concatenation, and partial traces are depicted by contracting the corresponding legs. See Appendix~\ref{OTOC-LE_relation_appendix} for further details.}
    \label{otoc_diagram}
\end{figure}

With this notation, we derive the OTOC--R\'enyi-entropy relation. For concreteness, we consider the infinite-temperature OTOC
\begin{equation}
F(t)=\Tr\!\left[\hat{R}_B^{\dagger}(t)\,\hat{W}^{\dagger}\,\hat{R}_B(t)\,\hat{W}\right],
\label{OTOC_infinite_def}
\end{equation}
shown diagrammatically in Fig.~\ref{otoc_diagram}. Here $\hat R_B$ is a unitary supported on subsystem $B$, and
\begin{equation}
\hat{R}_B(t)=\hat{U}^{\dagger}(t)\,\hat{R}_B\,\hat{U}(t)
\end{equation}
is its Heisenberg evolution under $\hat U(t)$.

\begin{figure}[ht] 
		\centering \includegraphics[width=0.85\textwidth]{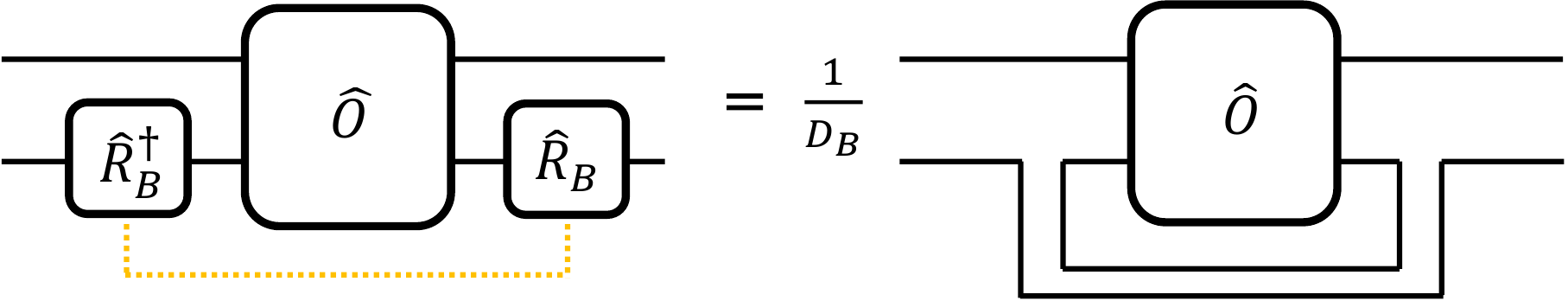} 
		\caption{The diagrammatic representation of the  Haar random integral formula Eq.~\eqref{Haar-random}. The orange dashed line in the left figure represents taking the Haar random average of the operator $\hat{R}_B$ defined on subsystem B. In the right figure, connecting the input and output legs of the subsystem $B$ corresponds to taking its partial trace.}
        \label{Haar_fig}
	\end{figure}

\begin{figure}[ht] 
		\centering \includegraphics[width=0.8\textwidth]{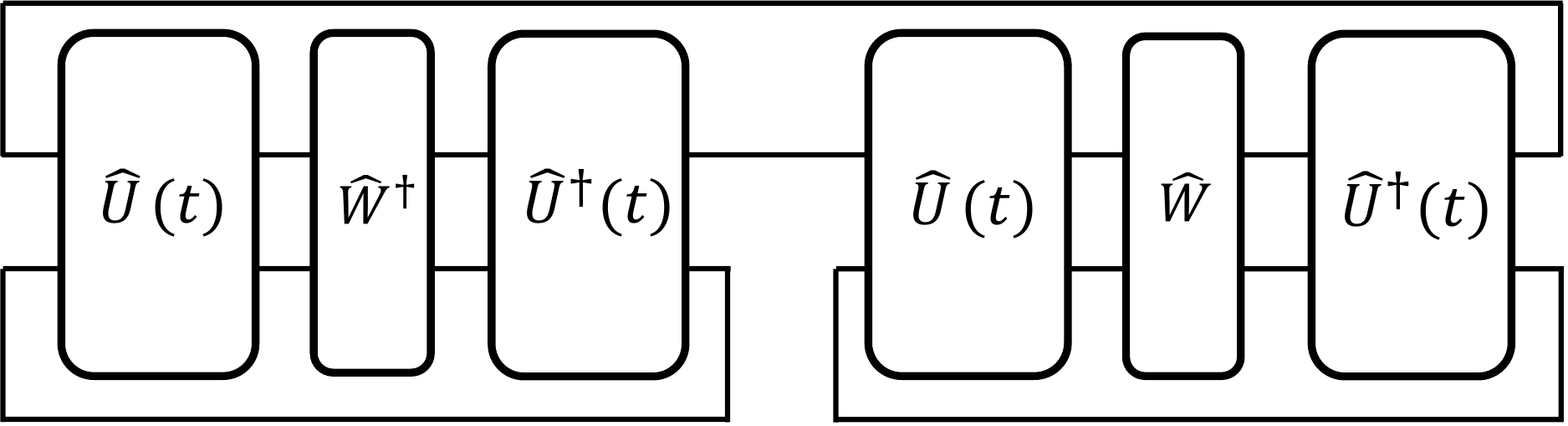}
		\caption{The diagram representation of the average of OTOC over $\hat{R}_B$ as defined in the Eq.~\eqref{average_OTOC_A} after using cyclic property of the trace.}
		\label{otoc_average2}
\end{figure}

Next, we consider the average OTOC by performing Haar random averaging over the operator $\hat{R}_B$ on subsystem B
\begin{equation}
	\overline{F(t)}=\int d\hat{R}_B   \text{Tr} \left[ \hat{R}_B^{\dagger}(t)\hat{W}^{\dagger}\hat{R}_B(t)\hat{W}\right].
 \label{average_OTOC_A}
\end{equation} 
We use the Haar random integral formula,\begin{equation}
	\int d\hat{R}_B \hat{R}_B^{\dagger} \hat{O} \hat{R}_B = \frac{1}{D_B}  \text{Tr}_B(\hat{O}) \otimes \hat{\ident}_B.
 \label{Haar-random}
\end{equation} 
This formula is depicted in Fig.~\ref{Haar_fig}.
\begin{figure}[ht] 
		\centering \includegraphics[width=0.8\textwidth]{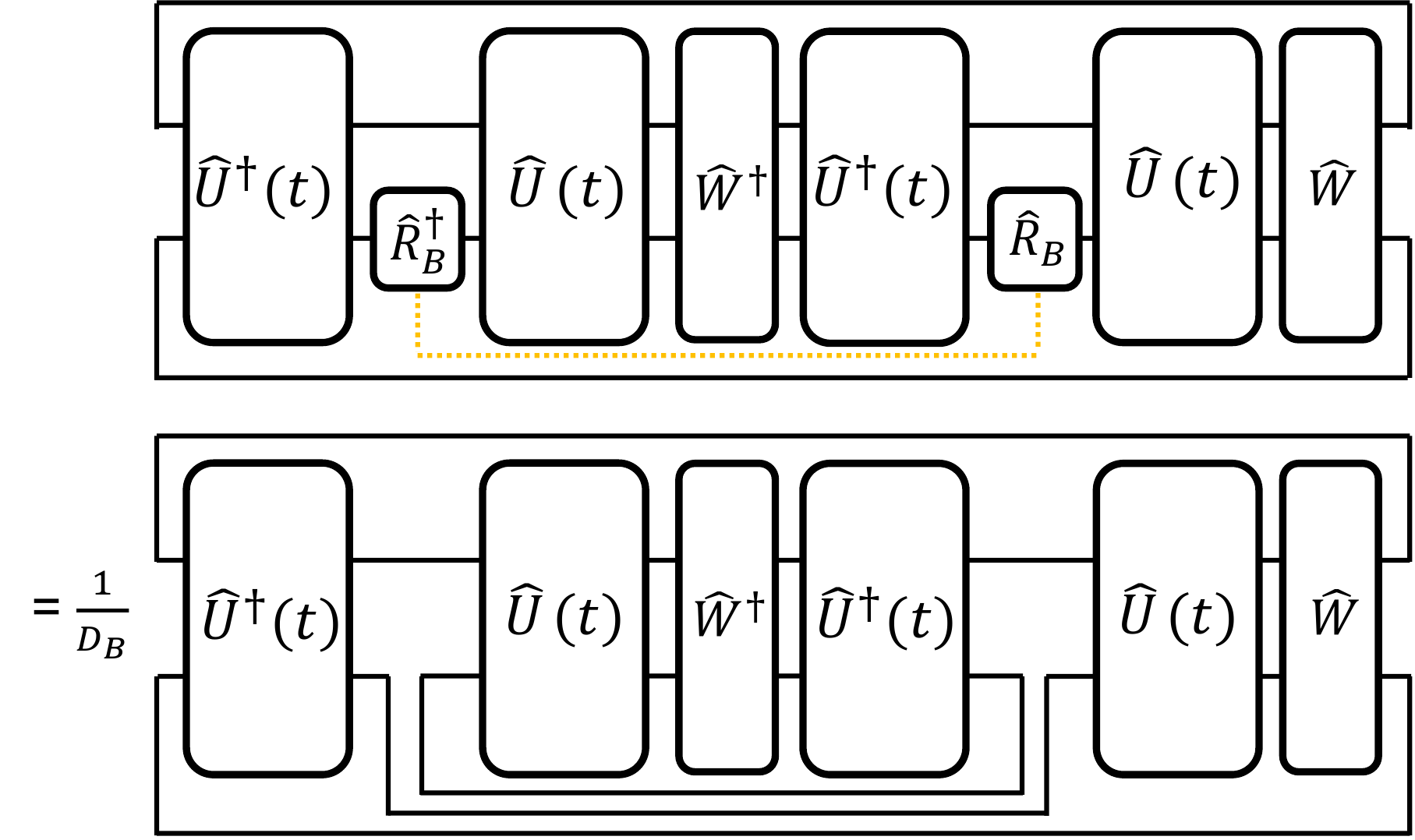} 
		\caption{The diagrammatic representation of the average of OTOC over $\hat{R}_B$ as defined in the Eq.~\eqref{average_OTOC_A}.}
		\label{otoc_average1}
\end{figure}
Then we have
\begin{equation}
    \begin{split}
        \int d\hat{R}_B \text{Tr}  \left[ \hat{R}_B^{\dagger}(t)\hat{W}^{\dagger}\hat{R}_B(t)\hat{W}  \right]
        &= \frac{1}{D_B}\text{Tr}  \left[ \text{Tr}_B[\hat{U}(t)\hat{W}^{\dagger}\hat{U}^{\dagger}(t)]\otimes \hat{\ident}_B \hat{U}(t)\hat{W}\hat{U}^{\dagger}(t) \right]\\
        &= \frac{1}{D_B}\text{Tr}_A  \left[ \text{Tr}_B[\hat{U}(t)\hat{W}^{\dagger}\hat{U}^{\dagger}(t)]\text{Tr}_B[\hat{U}(t)\hat{W}\hat{U}^{\dagger}(t)] \right]\\
        &=\text{Tr}_A \left[\hat{\rho}_A^2(t)\right].
    \end{split}
\end{equation} 
In the last step, we set $\hat{W}=\sqrt{D_B}\hat{\rho}(0)$, which gives
\begin{equation}
	\text{Tr}_B\left[\hat{U}(t)\hat{W}^{\dagger}\hat{U}^{\dagger}(t)\right]=\text{Tr}_B\left[\hat{U}(t)\hat{W}\hat{U}^{\dagger}(t)\right]=\sqrt{D_B}\text{Tr}_B\left[\hat{U}(t)\hat{\rho}(0)\hat{U}^{\dagger}(t)\right]=\sqrt{D_B}\hat{\rho}_A(t).
\end{equation}
Combining the above two equations, we have
\begin{equation}
    \int d\hat{R}_B \text{Tr}  \left[ \hat{R}_B^{\dagger}(t)\hat{W}^{\dagger}\hat{R}_B(t)\hat{W}  \right]=\text{Tr}_A \left[\hat{\rho}_A^2(t)\right].
\end{equation} 
The diagrammatic representation of the average OTOC in Eq.~\eqref{average_OTOC_A} is shown in Fig.~\ref{otoc_average1}, where the orange dashed line represents the Haar random average of the operator $\hat{R}_B$. Then, we use the cyclic property of the trace, the diagrammatic representation of the average OTOC can be further represented as in Fig.~\ref{otoc_average2}.

\begin{figure}[ht] 
		\centering \includegraphics[width=0.85\textwidth]{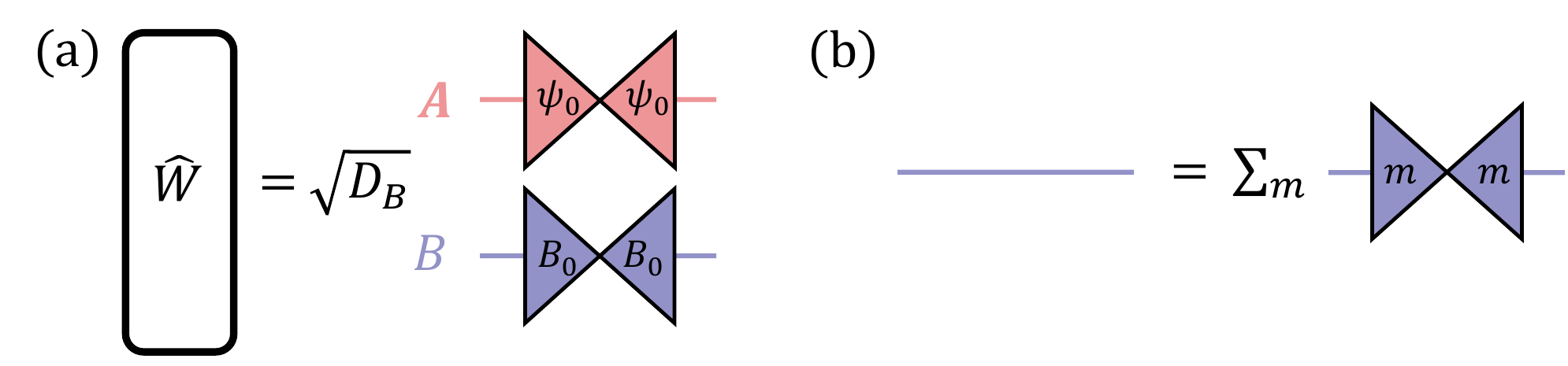} 
		\caption{(a) The diagrammatic representation of the initial density matrix $\hat{\rho}(0) = |\psi_0\rangle_{AA} \langle \psi_0| \otimes |B_0\rangle_{BB} \langle B_0|$ is shown. A ket-state $|\psi_0\rangle$ is represented by a triangle with a left leg, while a bra-state $\langle \psi_0|$ is represented by a triangle with a right leg.  (b) The diagrammatic representation of the identity operator is simply shown as a line. It can also be expressed as $\hat{\ident} = \sum_m |m\rangle \langle m|$, where $\{|m\rangle\}$ forms a complete orthonormal basis of the corresponding Hilbert space.}
		\label{initial_rho_fig}
\end{figure}

\begin{figure}[ht] 
		\centering \includegraphics[width=0.85\textwidth]{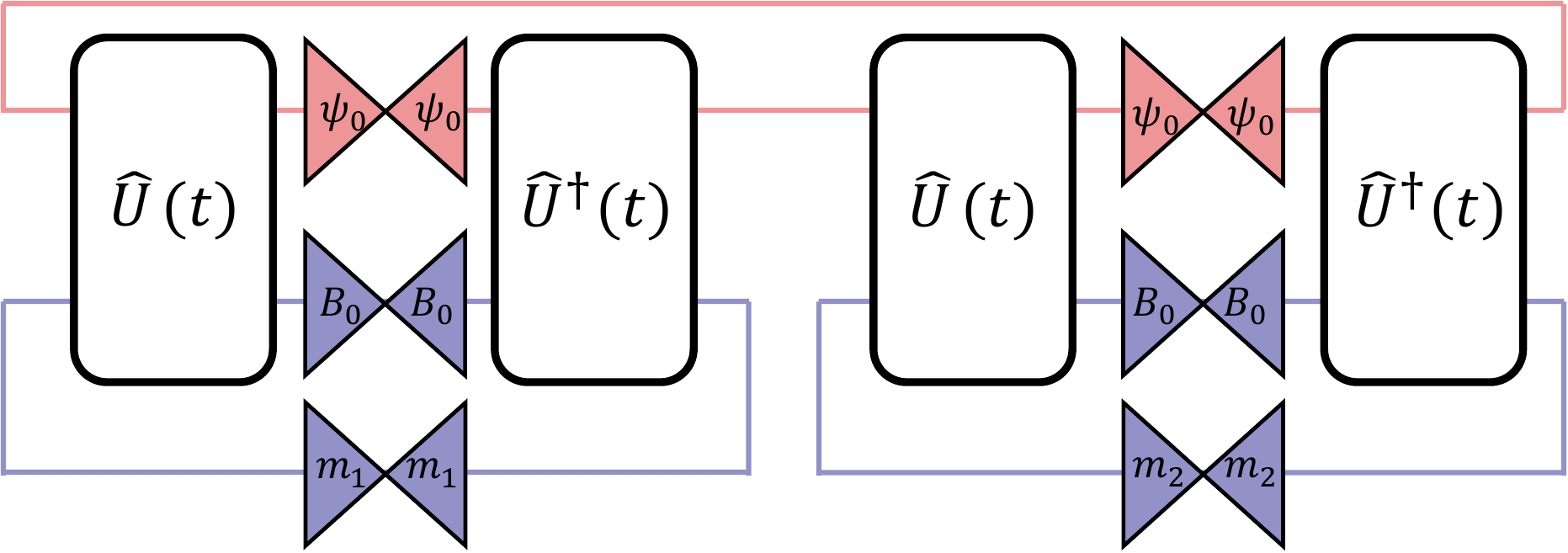} 
		\caption{The diagram representation of average OTOC defined in Eq.~\eqref{average_OTOC_A} with $\hat{W}=\sqrt{D_B}\hat{\rho}(0)=\sqrt{D_B}|\psi_0\rangle_{AA}\langle \psi_0|\otimes |B_0\rangle_{BB}\langle B_0|$. Here, we use pink to represent subsystem A and purple to represent subsystem B to help the Reader distinguish between the two subsystems.}
		\label{purity_fig}
\end{figure}

Thus, we obtain a general relation between the OTOC and the purity,
\begin{equation}
	\text{Tr}[\hat{\rho}_A^2(t)]=\frac{1}{D_B}\int d\hat{R}_B   \text{Tr} \left[ \hat{R}_B^{\dagger}(t)\hat{W}^{\dagger}\hat{R}_B(t)\hat{W}\right].
\end{equation} 
Accordingly, we derive the OTOC-R\'enyi entropy relation \cite{fanOutofTimeOrderCorrelationManyBody2017}:
\begin{equation}
	e^{-S_A^{(2)}}=\frac{1}{D_B}\int d\hat{R}_B   \text{Tr} \left[ \hat{R}_B^{\dagger}(t)\hat{W}^{\dagger}\hat{R}_B(t)\hat{W}\right].
\end{equation}
with $\hat{W}=\sqrt{D_B}\hat{\rho}(0)$. Recall the R\'enyi entropy--basis-resolved ETP relation we have derived in Eq.~\eqref{RE-LE_relation_equation}. Combining these two relations, we have the OTOC--basis-resolved ETP relation
\begin{equation}
	\frac{1}{D_B}\int d\hat{R}_B   \text{Tr} \left[ \hat{R}_B^{\dagger}(t)\hat{W}^{\dagger}\hat{R}_B(t)\hat{W}\right]=\sum_{m_1,m_2=1}^{D_B} M(t, m_1, m_2).
    \label{OTOC-LE_equation}
\end{equation} 
Here, $M(t, m_1, m_2)$ is the basis-resolved ETP defined in Eq.~\eqref{LE_projeted_def}, and $\hat{W}$ in the OTOC is chosen as $\hat{W} = \sqrt{D_B}\hat{\rho}(0) = \sqrt{D_B}|\psi_0\rangle_{AA} \langle \psi_0| \otimes |B_0\rangle_{BB} \langle B_0|$, as shown in Fig.~\ref{initial_rho_fig} (a). Also, the identity matrix is represented in Fig.~\ref{initial_rho_fig} (b).
Choosing the initial density matrix as the operator $\hat{W}$ in the average OTOC, the average OTOC can be further represented using the diagrammatic technique as in Fig.~\ref{purity_fig}.
This diagram already illustrates how the left-hand side of the Eq.~\eqref{OTOC-LE_equation} can be measured as the sum of basis-resolved ETP. To further aid the Reader, we provide a guided figure, Fig.~\ref{RE_LE_fig} in the Appendix \ref{OTOC-LE_relation_appendix}, to make this interpretation clearer.

We have thus derived the OTOC--basis-resolved ETP relation Eq.~\eqref{OTOC-LE_equation}, which states that the average OTOC, taken over the Haar random ensemble\footnote{Strictly speaking, a  unitary $1-$design suffices.} of operators defined on the traced-out part of the subsystem $B$, can be directly expressed as the sum of basis-resolved ETP.

\section{Applications}
\label{application}
In this section, we discuss two experimental applications of our R\'enyi entropy measurement protocol. First, we outline a superconducting circuit implementation that measures the second R\'enyi entropy via the basis-resolved ETP, illustrated with a three-qubit example. Second, we describe a cavity-QED application, where the same protocol enables the construction of holographic Hamiltonians.

\subsubsection*{Application to the superconducting circuit platform}
\label{circuit_protocol}
\begin{figure}[t] 
    \centering \includegraphics[width=0.9\textwidth]{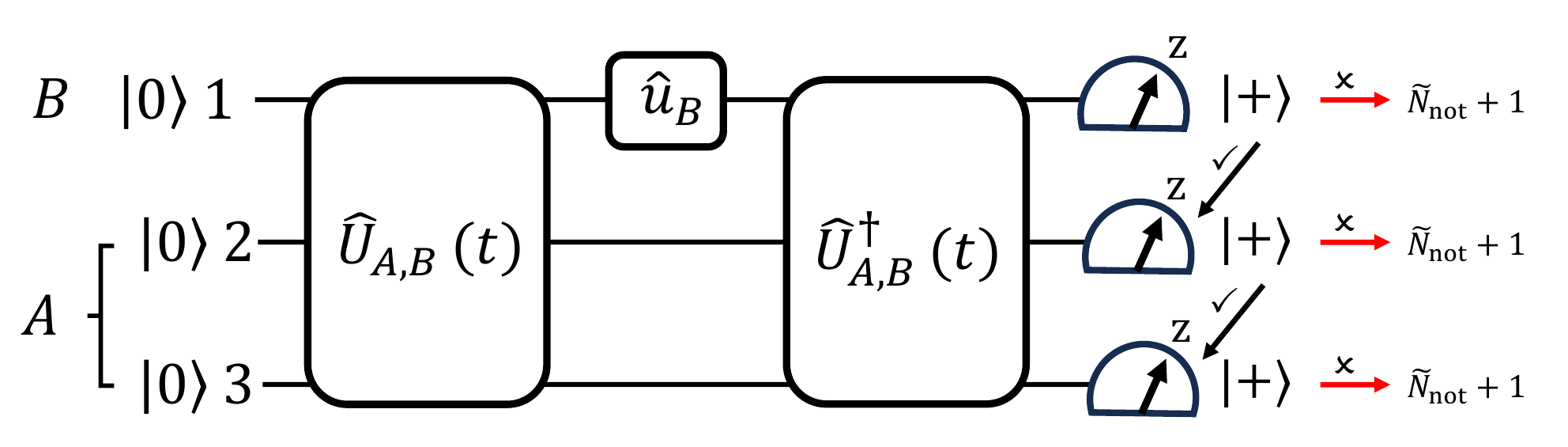} 
    \caption{ The protocol for measuring the second R\'enyi entropy using the basis-resolved ETP measurement protocol begins with the initial state $|0,0,0\rangle$ in a 3-qubit superconducting circuit.}
\label{circuit_proposal}
\end{figure}

We present a minimal proposal for measuring the second R\'enyi entropy on a superconducting circuit platform. Such platforms naturally enable time-reversal operations, which are essential for implementing Loschmidt echo type of experiments~\cite{Braum_ller_2021}. For simplicity, we consider a three-qubit system. Qubit~1 is assigned to subsystem~$B$ ($N_B=1$), while qubits~2 and~3 form subsystem~$A$ ($N_A=2$). Using the simplified protocol shown in Fig.~\ref{Renyi2_protocol_random_simplified}, only a single copy of subsystems $A$ and $B$ is required; thus, a total of three qubits ($N_A+N_B=3$) suffices to measure the second R\'enyi entropy. The general measurement scheme described in Sec.~\ref{measure_Renyi_entropy_subsection} can be directly applied to this three-qubit system. The corresponding circuit implementation is shown in Fig.~\ref{circuit_proposal}. The second R\'enyi entropy can be computed as in Eq.~\eqref{Renyi_entropy_protocol_equation}.

\subsubsection*{Application to holographic cQED platforms}
\label{Holographic_protocol_appendix}

The protocol introduced in Sec.~\ref{RE-LE_relation} enables efficient measurements of two scrambling-related observables in many-body systems. Scrambling is central to holography: the dual black-hole picture motivates holographic models that saturate the fast-scrambling conjecture~\cite{Sekino_2008}, which bounds the scrambling time to grow at most logarithmically with system size. Measuring scrambling in such models is therefore of particular interest, as it may provide indirect probes of black-hole physics. This naturally raises the question of whether our protocol can be implemented on platforms that realize holographic Hamiltonians. In this subsection, we show that cQED platforms provide one pertinent example. Recently, advances have been made for models with both random couplings \cite{Uhrich_2023, Baumgartner_2024} and fixed couplings \cite{Bentsen_2019, Periwal_2021, Davis_2019}. Time reversal in models with random couplings is a-priori hard to realize, as precise control over each individual coupling is required to reverse the sign of every term. We, therefore, focus on models with fixed couplings.

An elegant method for constructing non-local couplings with high controllability was proposed in \cite{Periwal_2021}. This approach employs an atomic lattice, where each site contains either a single atom or an atomic cloud exposed to a magnetic field perpendicular to the cavity axis. Interactions between different sites are mediated by double Raman scattering of photons. Adiabatic elimination of the photons yields an effective Hamiltonian of the form \cite{Davis_2019, Periwal_2021}
\begin{equation*}
    \hat{H}_{\text{eff}}=\sum_{i=1}^N \chi_{ii}\hat{T}_{ii}+\sum_{i,j=1}^N \Big[\chi_{ij}\hat{T}_{ij}e^{-i\omega_{ij}t} + \chi_{ij}^*\hat{T}^\dagger_{ij}e^{i\omega_{ij}t}\Big]
\end{equation*}
with transition operator $\hat{T}_{ij}=\hat{L}_i^+\hat{L}_j^-$, where $\hat{L}^{\pm}_i$ are angular momentum ladder operators and $i$ denotes the position in the spin chain of length $N$. Moreover, $\omega_{ij}=\omega_j-\omega_i$ corresponds to the energy difference between the Zeeman-splittings of sites $i$ and $j$. A linear, non-constant B-field allows for the elimination of all cross-couplings in the lattice. By engineering additional sidebands in the drive laser at separation $\pm\omega_{ab}$, selective cross-couplings can be reinstalled. The amplitude and sign of $\chi_{ij}$ can as well be controlled by the laser. \\
\begin{figure}[t!] 
		\centering \includegraphics[width=0.7\textwidth]{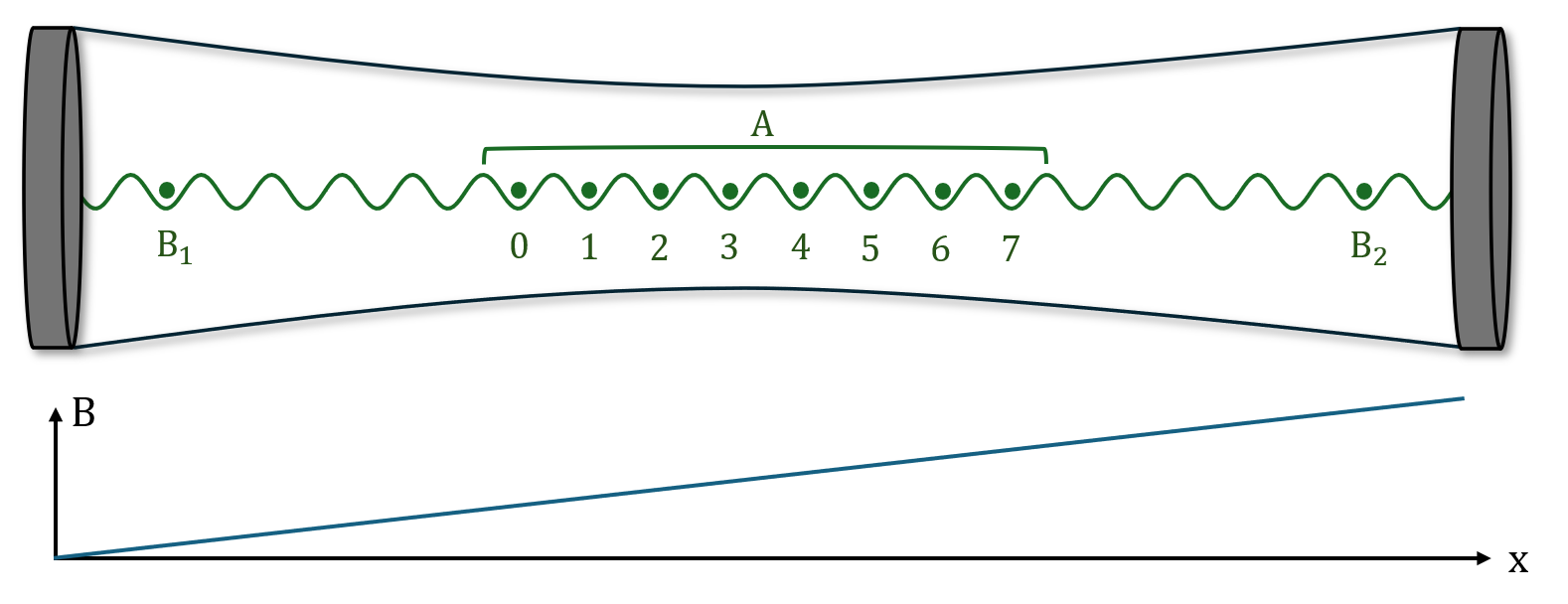} 
		\caption{Sketch illustrating the implementation of the 2-adic model with $N=2^3$. The subsystems $B_1$ and $B_2$ are given by two additional sites located to the left and right of the chain. The energy separation from $A$ must be greater than $\Delta\omega_{\text{max}}$ of the sidebands to prevent unwanted cross-couplings. Here, this requires $|B_1-0|>7$ and $|B_2-7|>7$.}
		\label{holographic_cavity_setup_fig}
\end{figure}
To implement our protocol, the lattice must be divided into three parts: the main system $A$ and two bath systems, $B_1$ and $B_2$. The baths must have equal size and they have to couple in the same way to $A$. Both can be naturally achieved with the setup, as the sidebands of the laser provide full control over the couplings. It is important to note that, in general, $A$, $B_1$ and $B_2$ should not be adjacent in the lattice. This is because constructing specific couplings in $A$ can lead to unavoidable cross-couplings with $B_1$ and $B_2$, thereby distorting the intended model. Thus, it is necessary to separate the systems far enough that those cross-couplings can be avoided. That means, when creating the atomic lattice, an additional step is necessary where the sites between the systems have to be emptied\footnote{It has been reported in \cite{Karski_2010, Kuhr_2003} that this can be done by using the so-called \textit{push-out} technique. Since the Zeeman splittings of the couplings are, by design, different at each site, this method should be suited for implementation here.}. Finally, a key requirement for implementing the protocol is time reversal. As noted earlier, the ability to modify the phase of each coupling allows for sign inversion, making time reversal naturally achievable. 

We conclude this discussion by considering a specific implementation of non-local interactions that realizes a truncated version of p-adic AdS/CFT \cite{Gubser_2017}, thereby providing a concrete example of a holographic model. The idea has been proposed in \cite{Bentsen_2019, Periwal_2021} and is based on the couplings
\begin{equation*}
    \chi_{ij}=\begin{cases} 
      |i-j|^s & |i-j|=p^n, \quad p\in\mathbb{P} \\
      0 ,& \text{else} 
   \end{cases}
\end{equation*}
where $s$ is a parameter that interpolates between a non-local, $p$-adic geometry ($s>0$) and a local Archimedean geometry ($s<0$). For $s=0$, the underlying geometry is a hypercube and thus naturally supports logarithmic scrambling and the possibility to saturate the fast scrambling conjecture \cite{Bentsen_2019}. Furthermore, for $s\geq 0$, the model exhibits a dual geometry given by the Bruhat-Tits tree. Note that the geometry is constructed with periodic boundary conditions, meaning $|0-N|=1$ for a chain with $N$ sites.

As a concrete example, consider the 2-adic model with $N=2^3$. The laser has three different sidebands corresponding to $\omega_{n(n+1)}$, $\omega_{n(n+2)}$, $\omega_{n(n+4)}$ and $\omega_{n(n+7)}$, where the latter implements the periodic boundary conditions. 
The bath systems $B_1$ and $B_2$ each consist of one additional site, with energy separation $\omega_{B_1 0}>\omega_{n(n+7)}$ and $\omega_{B_2 7}>\omega_{n(n+7)}$ to avoid unwanted couplings. They can be coupled to system $A$ (e.g., both to sites $0$ and $7$) via additional laser sidebands. The duration of unitary evolution can be controlled by switching the lasers on and off. Finally, using these techniques, we can implement the protocols described in Sections \ref{sec:general_proposal} and \ref{sec:random_unitary_approach}. Figure \ref{holographic_cavity_setup_fig} provides a schematic of the system for the 2-adic case.

\section{Summary \& Discussion}
\label{summary}

\begin{figure}[t] 
    \centering \includegraphics[width=0.8\textwidth]{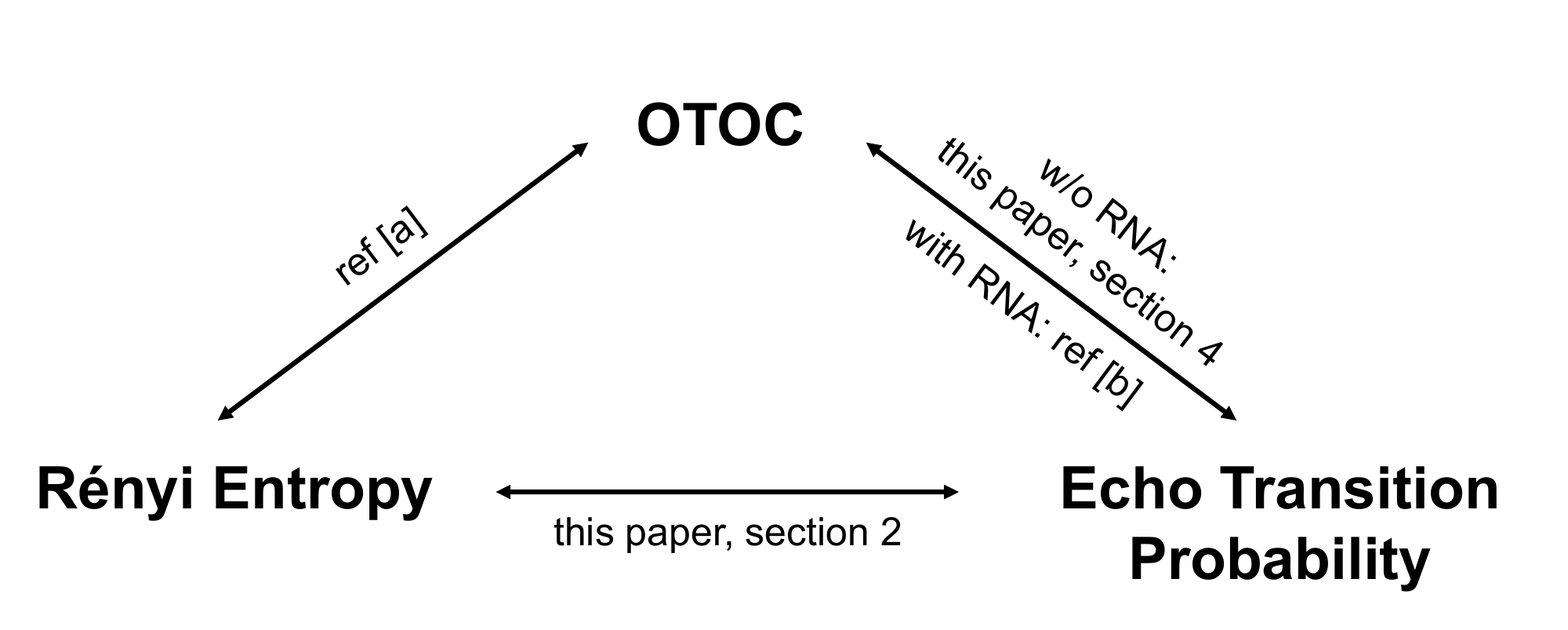} 
    \caption{
    The triangular relation among the OTOC, R\'enyi entropy, and echo transition probability (ETP) is illustrated. The Loschmidt echo is a special case of the ETP, corresponding to the choice where the target state coincides with the initial state. The R\'enyi-entropy--ETP relation is presented in Sec.~\ref{RE-LE_relation}. The relation between R\'enyi entropy and OTOC was derived in (a) \cite{fanOutofTimeOrderCorrelationManyBody2017}, and the relation between OTOC and Loschmidt echo was obtained in (b) \cite{yanInformationScramblingLoschmidt2020}, where a random-noise average (RNA) is required. In addition, in Sec.~\ref{OTOC-LE_relation_main} we establish the OTOC--ETP relation without requiring RNA.}
\label{triangle_fig}
\end{figure}

In this paper, we derived a mathematical relation between the R\'enyi entropy and the LE. We found that the exponential of the second R\'enyi entropy, which is also known as the quantum purity, can be expressed as the sum of the basis-resolved ETP, which we defined in this paper. Based on this, we designed a protocol for measuring R\'enyi entropy using existing LE measurement protocols. Our results thus provide a further vertex in a triangle of relations between echo transition probability, OTOCs, and R\'enyi entropies, as summarized in Fig. \ref{triangle_fig}.
Furthermore, we provided a diagrammatic proof of the OTOC-basis-resolved ETP relation, notably avoiding the need for a random noise ensemble average by combining the R\'enyi entropy-LE relation with the known OTOC-R\'enyi entropy relation.  We presented two examples demonstrating that our protocol can be implemented on existing platforms, firstly using superconducting circuits, where direct time reversal is possible, and secondly, using cavity QED systems, which enable the construction of holographic Hamiltonians. Additionally, in Appendix \ref{measure_RE_without_time_reversal}, we give a method for measuring R\'enyi entropy without requiring time reversal, using the technique of randomized measurements, which may make R\'enyi entropy measurement more accessible on experimental platforms where direct time reversal remains challenging. Furthermore, in Appendix \ref{n-th_Renyi_appendix}, we present a method to measure the $n-$th R\'enyi entropy $(n \geq 2)$ using the basis-resolved ETP protocol and derive its upper and lower bounds in terms of basis-resolved ETP.

Our protocol for measuring the second R\'enyi entropy of subsystem $A$ requires only a single copy of subsystem $A$ and a single subsystem $B$. Each round consists of a forward joint evolution of $A$ and $B$, followed by either (i) resetting $B$ to the given initial state or (ii) applying a random unitary to randomize $B$, and then a backward evolution of $A$ and $B$. This approach is similar to the protocol used for measuring the LE and is directly measurable on experimental platforms where such echo experiments are realizable, such as the superconducting qubits \cite{Braum_ller_2021}.

Our method is more resource-efficient than previous protocols for measuring the second R\'enyi entropy, which involves preparing two copies of the entire system \cite{Islam_2015,Kaufman_2016,Linke_2018,PhysRevX.9.031013}. Moreover, in our protocol, measurements are limited to a smaller part of the system (subsystem $B$) if the measurement on $B$ does not yield the initial value. This significantly reduces resource requirements, particularly in scenarios where the time evolution is long and chaotic, the quantum purity is low (indicating a low probability for the final state of $B_1$ to match the given specific state), or subsystem $A$ is very large. 

Our protocol for measuring the entanglement entropy is not restricted to non-inter\-acting systems, where the entanglement entropy can be derived from the correlation matrix obtained by measuring the system's two-point functions \cite{Lin_2024}. Furthermore, the measurement of the larger subsystem A can be optimized to reduce the number of qubits that need to be accurately measured by considering the correlations in the final state or utilizing classical shadow tomography, which allows for constructing an approximate classical description of a quantum state using only a few measurements \cite{aaronson2018shadowtomographyquantumstates, Huang_2020, Huang_2022, PhysRevResearch.5.023027}. This is a promising avenue to consider, and we leave the development of improved protocols for future exploration. Also, in our protocol for measuring R\'enyi entropy, the complexity of implementing time reversal or using randomized measurements as a substitute for direct time reversal could be exponentially high in the worst case \cite{Nielsen_Chuang_2010, PRXQuantum.2.030316, Aaronson:2022flk}. It would be very interesting to investigate this further in future work. In the meantime, we do not consider this a significant obstacle for the system sizes that can now be realized on near-term quantum computers.

From a theoretical perspective, in the study of quantum chaos and quantum information scrambling, researchers have explored the relations among key quantities such as the OTOC, R\'enyi entropy, Loschmidt echo (LE), spectral form factor, etc \cite{Cotler_2017, de_Mello_Koch_2019, Romero_Berm_dez_2019, Kudler_Flam_2020, toniolo2025dynamicalalpharenyientropieslocal}. For example, the OTOC can be expressed as the thermal average of the LE \cite{yanInformationScramblingLoschmidt2020}, while the R\'enyi entropy can be written as the random average of the OTOC \cite{fanOutofTimeOrderCorrelationManyBody2017}. In this paper, we establish a direct connection between the LE (special case of ETP) and R\'enyi entropy. By combining our findings with previous results, we provide a triangular relationship among the OTOC, LE, and R\'enyi entropy, with all pairwise relationships between these three quantities fully derived. Consequently, in experiments, measuring any one of these quantities allows researchers to infer information about the other two. 

Moreover, based on the R\'enyi entropy–ETP relation we have derived, there is no need to rely on a random noise ensemble to represent the time evolution of the reduced density matrix, as was necessary in previous works discussing the OTOC–LE relation \cite{yanInformationScramblingLoschmidt2020}. Furthermore, the generalization of this triangular relation in open quantum systems is an intriguing question to explore whether it still holds or if dissipation alters its form \cite{Zhou_2021, zhou2024generalizedloschmidtechoinformation, Zhou_2024}. This topic is particularly relevant to real experiments, where dissipation is nearly unavoidable.

Finally, we comment on how our protocol compares with classical shadows in terms of resource scaling. Here ``efficiency'' means \emph{hardware and measurement-setting efficiency}, not an information-theoretic reduction in sample complexity. Classical shadows estimate quantities such as $\Tr(\hat{\rho}_A^2)$ from randomized measurement ensembles combined with classical post-processing, and are particularly useful when one wishes to predict \emph{many} observables from the same data~\cite{Huang_2020}. In contrast, our basis-resolved ETP protocol is tailored to echo-capable platforms: it uses only forward--backward evolution, a single reusable ancilla $B$ (reset between the two evolutions), fixed-basis $\sigma_z$ readout, and minimal post-processing (a failure counter), thereby avoiding deep randomizing circuits and large measurement-setting overhead. We do \emph{not} claim fewer shots for purity estimation: Appendix~\ref{app:purity_sample_complexity} shows that achieving relative error $\eta$ requires $N_{\mathrm{cycle}}\gtrsim(\eta^2 x)^{-1}$ and hence $N_{\mathrm{shots}}=D_B N_{\mathrm{cycle}}\gtrsim D_B/(\eta^2 x)$ for $x=\Tr(\hat{\rho}_A^2)$, which becomes exponential when $x$ is exponentially small. This rare-event scaling similarly limits randomized-measurement/shadow-based purity estimators~\cite{BrydgesEtAl2019ProbingRenyi,PhysRevLett.120.050406}. Thus, the two approaches are complementary: shadows trade randomized control and post-processing for broad multi-observable access, whereas the basis-resolved ETP protocol minimizes control complexity and measurement settings to target a specific nonlinear functional via echo dynamics.

\section*{Acknowledgements}
It is a pleasure to thank Rahel Baumgartner, Jordan Cotler, Netta Engelhardt, and Pietro Pelliconi for discussions. This work has received funding through the Swiss Quantum Initiative awarded by the State Secretariat for Economic Affairs, under the grant "HoloGraph". This research is supported in part by the Fonds National Suisse de la Recherche Scientifique (Schweizerischer Nationalfonds zur Förderung der wissenschaftlichen For\-schung) through the Project Grant 200021\_215300 and the NCCR51NF40-141869 The Mathematics of Physics (SwissMAP). JS thanks Harvard University for support through a Bershadsky Distinguished Visiting Fellow award.


\bibliographystyle{quantum}
\bibliography{ref.bib}

\appendix

\section{The measurement of R\'enyi entropy without time reversal}\label{measure_RE_without_time_reversal}
The measurement of the R\'enyi entropy through the measurement of the LE involves time reversal, as it requires performing backward time evolution. Time reversal can be realized by exactly reversing the sign of the Hamiltonian. To achieve this, one must fine-tune the experimental parameters to precisely reverse the sign of every term in the Hamiltonian. Alternatively, it can be implemented by coupling the system to an ancilla \cite{PhysRevLett.119.040501}. However, in some experimental platforms, direct time reversal may be challenging using currently available technologies. Alternatively it can be avoided by using randomized measurements \cite{PhysRevLett.108.110503,Vermersch_2019,Elben_2019,PhysRevLett.120.050406,Joshi_2020, Elben_2022}, by relying on the concept of unitary designs. In this Appendix we show how to apply this idea to the case at hand.

The key idea behind using randomized measurements to eliminate the need for time reversal is to apply the formula for unitary designs—specifically, the unitary $2$-design as an intermediate step in the protocol. One effectively substitutes the concept of temporal correlation after time inversion by that of correlation with respect to an ensemble of measurements. Mathematically speaking, this approach replaces a single trace quantity (which requires time reversal) with the product of two single-trace quantities, both of which are evaluated without time reversal, under a random average.

More precisely, using the formula for a random unitary $\hat{u}$ that satisfies the properties of a unitary $2$-design, one obtains the following relation for two general operators $\hat{R}$ and $\hat{S}$ \cite{Vermersch_2019}:  
\begin{equation}
    \begin{split}
        \overline{\langle \hat{R} \rangle_u \langle \hat{S}\rangle_u}=&\frac{1}{D_H^2-1}\bigl[\text{Tr}(\hat{\rho}_0)^2\text{Tr}(\hat{R})\text{Tr}(\hat{S})+\text{Tr}(\hat{\rho}_0^2)\text{Tr}(\hat{R}\hat{S})\bigr]\\
        &-\frac{1}{D_H(D_H^2-1)}\bigl[\text{Tr}(\hat{\rho}_0)^2\text{Tr}(\hat{R}\hat{S})+\text{Tr}(\hat{\rho}_0^2)\text{Tr}(\hat{R})\text{Tr}(\hat{S})\bigr].
    \end{split}
\end{equation}   
Here, $\langle \hat{R} \rangle_u = \text{Tr}(\hat{u}^{\dagger} \hat{\rho}_0 \hat{u} \hat{R})$, the overline denotes the random average over $\hat{u}$ with respect to the Haar measure, and $D_H$ is the Hilbert space dimension of the total system.

In the above formula, by setting $\hat{R} = \hat{S} = \hat{\rho}(t)$ and using $\text{Tr}[\hat{\rho}(t)] = 1$, we obtain:
\begin{equation}
        \overline{\langle \hat{\rho}(t) \rangle_u \langle \hat{\rho}(t)\rangle_u}=\frac{1}{D_H^2-1}\bigl[1+\text{Tr}(\hat{\rho}_0^2)\text{Tr}(\hat{\rho}^2(t))\bigr]-\frac{1}{D_H(D_H^2-1)}\bigl[\text{Tr}(\hat{\rho}^2(t))+\text{Tr}(\hat{\rho}_0^2)\bigr].
    \label{purity_random_measurement}
\end{equation}
If we measure the quantity on the left-hand side at time $t=0$, then we have
\begin{equation}
    \overline{\langle \hat{\rho}(t) \rangle_u \langle \hat{\rho}(t)\rangle_u}\Bigr|_{t=0}=\frac{1}{D_H^2-1}\left\{1+[\text{Tr}(\hat{\rho}_0^2)]^2\right\} -\frac{1}{D_H(D_H^2-1)}\bigl[2\text{Tr}(\hat{\rho}_0^2)\bigr].
    \label{purity_initial}
\end{equation}

Using Eq.~\eqref{purity_initial}, one can solve for the value of $\text{Tr}(\hat{\rho}_0^2)$, and by combining it with Eq.~\eqref{purity_random_measurement}, one can determine the value of $\text{Tr}(\hat{\rho}^2(t))$.

The experimental protocol of measuring the left-hand side of Eq.~\eqref{purity_random_measurement} is as follows:

(i) Prepare the initial density matrix $\hat{\rho}_0$.

(ii.a) In the first experiment, evolve the system in time with $\hat{U}(t)$ and then apply a global random unitary $\hat{u}$ to it to obtain $\hat{u}\hat{U}(t)\hat{\rho}_0 \hat{U}^{\dagger}(t)\hat{u}^{\dagger}$. Then, measure the probability that the final state returns to the initial state. Repeat steps (i) and (ii.a) with the same random unitary $\hat{u}$ to measure $\langle \hat{\rho}(t) \rangle_u$.

(ii.b) In the second experiment, after step (i), we first apply a global random unitary $\hat{u}$ to the initial state without time evolution and then measure the probability of the final state returning to the initial state. Repeat steps (i) and (ii.b) with the same random unitary $\hat{u}$ to measure $\langle \hat{\rho}(0) \rangle_u$.

Finally, we repeat steps (i) and (ii) for different random unitaries. The purity at the initial time $t = 0$ can be obtained from the second experiment, as defined in Eq.~\eqref{purity_initial}, and is calculated from the statistical correlation $\overline{\langle \hat{\rho}(t) \rangle_u \langle \hat{\rho}(t) \rangle_u}\Bigr|_{t=0}$.

The purity at a general time $t$ can be determined from the first experiment using Eq.~\eqref{purity_random_measurement} and the initial purity value obtained from the second experiment using Eq.~\eqref{purity_initial}.

\section{Measuring the n-th R\'enyi entropy via basis-resolved ETP protocol}
\label{n-th_Renyi_appendix}

\label{nth_Renyi-bound}
Here, we consider the relation between the $n$-th R\'enyi entropy and the ETP. The $n$-th R\'enyi entropy is defined by
\begin{equation}
S_A^{(n)}=\frac{1}{1-n}\log\left[\text{Tr}_A (\hat{\rho}_A^n)\right].
\end{equation}
Below, we define $B^{\otimes n}=B_1\cup B_2\cup\cdots\cup B_n$. If the initial density matrix is
$\hat{\rho}(0) = \hat{\rho}_A^0 \otimes \hat{\rho}_{B}^0=|\psi\rangle_{AA} \langle \psi| \otimes |\phi \rangle_{BB} \langle \phi|$, we have
\begin{equation}
\label{n_Renyi_def}
\begin{aligned}
    &\text{Tr}_A (\hat{\rho}_A^n)\\
    =&\text{Tr}_A \left[ \text{Tr}_{1}(\hat{U}_1\hat{\rho}_A^0 \otimes \hat{\rho}_{1}^0\hat{U}^{\dagger}_{1})\text{Tr}_{2}(\hat{U}_{2}\hat{\rho}_A^0 \otimes \hat{\rho}_{2}^0\hat{U}^{\dagger}_{2}) \dots\text{Tr}_{j}( \hat{U}_{j}\hat{\rho}_A^0 \otimes \hat{\rho}_{j}^0\hat{U}^{\dagger}_{j} ) \dots \text{Tr}_{n}( \hat{U}_{n}\hat{\rho}_A^0 \otimes \hat{\rho}_{n}^0\hat{U}^{\dagger}_{n} ) \right] \\
    =&\text{Tr}_{A\cup B^{\otimes n}}  \left[(\hat{\rho}_A^0 \otimes \hat{\rho}_{1}^0\hat{U}^{\dagger}_{1} \hat{U}_2) (\hat{\rho}_A^0 \otimes \hat{\rho}_{2}^0 \hat{U}^{\dagger}_2 \hat{U}_3)\dots (\hat{\rho}_A^0 \otimes \hat{\rho}_{j}^0 \hat{U}^{\dagger}_{j} \hat{U}_{j+1}) \dots  (\hat{\rho}_A^0 \otimes \hat{\rho}_{n}^0 \hat{U}^{\dagger}_n \hat{U}_1) \right] \\
    =& \sum_{\alpha=1}^{n} \sum_{b^{i_{\alpha}}_1,b^{i_{\alpha}}_2,\dots,b^{i_{\alpha}}_n=1}^{D_B}\prod_{j=1}^n\langle \psi,b^{i_j}_1,b^{i_j}_2,\dots,b^{i_j}_n|     \hat{\rho}_{j}^0\hat{U}^{\dagger}_{j} \hat{U}_{[j+1]}       \hat{\rho}_{[j+1]}^0|\psi,b^{i_{[j+1]}}_1,b^{i_{[j+1]}}_2,\dots,b^{i_{[j+1]}}_n \rangle \, \\
    = &   \sum_{m_1,m_2,\dots ,m_n=1}^{D_B}\prod_{j=1}^n \langle \psi,\phi_j,m_{[j+1]}|   \hat{U}^{\dagger}_{j} \hat{U}_{[j+1]} |\psi,m_j,\phi_{[j+1]} \rangle. \\
\end{aligned}
\end{equation}
For simplicity, we denote $\hat{U}_{A\cup B_j}(t)$ as $\hat{U}_j$ and $\hat{\rho}_{B_j}$ as $\hat{\rho}_j$. The definition of $[j]$ is 
\begin{equation}
    [j]= \begin{cases}
        j, \ \ \ \ \ \ \ \ 1\leq j \leq n \\
        j-n, j > n. \\
        j+n, j <1. \\
    \end{cases}
\end{equation} 
Here, $ |b^{i_j}_1, b^{i_j}_2, \dots, b^{i_j}_n\rangle = |b^{i_j}_1\rangle \otimes |b^{i_j}_2\rangle \otimes \dots \otimes |b^{i_j}_n\rangle $ represents an $ n $-dimensional vector that forms a complete basis for $ B^{\otimes n} $. By inserting the identity $$ \hat{\ident}_{B^{\otimes n}} = |b^{i_j}_1, b^{i_j}_2, \dots, b^{i_j}_n\rangle \langle b^{i_j}_1, b^{i_j}_2, \dots, b^{i_j}_n| $$ between each pair of parentheses in the third line, we obtain the expression in the fourth line. We have inserted a total of $n$ independent $ n $-dimensional vectors, labeling their indices from $ i_1 $ to $i_n$. The upper index $ i_p $ in $ |b^{i_p}_q\rangle $ denotes that the vector occupies the $p-$th position in this sequence, while the lower index $q$ indicates its association with the basis of subsystem $ B_q $.

The square of each component $\langle \psi,\phi_j,m_{[j+1]}^j|   \hat{U}^{\dagger}_{j} \hat{U}_{[j+1]} |\psi,m^{[j+1]}_j,\phi_{[j+1]} \rangle $ in the above equation is a basis-resolved ETP. However, since only its norm can be directly measured in the basis-resolved ETP experimental protocol, but not its phase, it is not directly measurable using that protocol. One would need to design a way to measure the relative phase using the basis-resolved ETP protocol.

Below, we present a method for measuring the relative phase in experiments in  \ref{measure_phase}. Since measuring this phase directly is challenging—it requires preparing subsystem B in a superposition of two basis states forming a complete basis—we also consider upper and lower bounds, which may be easier to measure. Additionally, we derive these bounds for the $n^{\rm th}$ R\'enyi entropy and explore their relation to basis-resolved ETP in \ref{lower_bound_appendix} and \ref{upper_bound_appendix}.

\subsection{Method for measuring the relative phase}
\label{measure_phase}
When an arbitrary initial state of subsystem B can be prepared in an experiment, the relative phase between the two components in the above equation also becomes measurable. We consider the example in which one wants to measure the relative phase between  
\begin{equation*}
\langle \psi,\phi_1,m_2|   \hat{U}^{\dagger}_1 \hat{U}_{2} |\psi,m_1,\phi_{2} \rangle
\end{equation*}  
and  
\begin{equation*}
\langle \psi,\phi_1,m_2|   \hat{U}^{\dagger}_1 \hat{U}_{2} |\psi,m_1^{'},\phi_{2} \rangle.
\end{equation*}  
We define  
\begin{equation}
\frac{\langle \psi,\phi_1,m_2|   \hat{U}^{\dagger}_1 \hat{U}_2 |\psi,m_1,\phi_2 \rangle}{\langle \psi,\phi_j,m_2|   \hat{U}^{\dagger}_{1} \hat{U}_{2} |\psi,m_1^{'},\phi_{2} \rangle} = e^{-i\beta} \left| \frac{\langle \psi,\phi_1,m_2|   \hat{U}^{\dagger}_1 \hat{U}_2 |\psi,m_1,\phi_2 \rangle}{\langle \psi,\phi_1,m_2|   \hat{U}^{\dagger}_{1} \hat{U}_2 |\psi,m_1^{'},\phi_2 \rangle} \right|\,.
\end{equation}  
Then, to determine the relative phase $\beta$, we proceed as follows. We begin by preparing a specific initial state that is a superposition of the states $|m_1\rangle$ and $|m_1^{'}\rangle$ (assumed to be orthogonal for simplicity). We define  
\begin{equation}
|m_{1,\alpha}\rangle = \frac{1}{\sqrt{2}}(|m_1\rangle + e^{i\alpha} |m_1^{'}\rangle),
\end{equation}  
and measure the basis-resolved ETP,  
\begin{equation}
    M(t,m_{1,\alpha},m_2) = |\langle \psi,\phi_1,m_2|   \hat{U}^{\dagger}_1(t) \hat{U}_2(t) |\psi,m_{1,\alpha},\phi_2 \rangle|^2.
\end{equation}  
Since we have  
\begin{equation}  
    \begin{aligned}  
        M(t,m_{1,\alpha},m_2)  
        = & \frac{1}{2}\left[M(t,m_1,m_2) + M(t,m_1^{'},m_2) \right] \\
        &+ \sqrt{M(t,m_1,m_2)M(t,m_1^{'},m_2)}\cos(\alpha+\beta),  
    \end{aligned}  
\end{equation}  
and both basis-resolved ETP, $ M(t,m_1,m_2) $ and $ M(t,m_1^{'},m_2) $, are measurable, one can uniquely determine the phase $\beta$ by measuring two basis-resolved ETP $M(t,m_{1,\alpha},m_2)$ and $M(t,m_{1,\alpha^{'}},m_2)$ with different phases $\alpha$ and $\alpha^{'}$. By combining these measurements with $ M(t,m_1,m_2) $ and $ M(t,m_1^{'},m_2) $, one can solve for $\beta$ using the following two equations:  

\begin{equation}  
    \begin{aligned}
        &\cos(\alpha+\beta) \\
        &= \frac{M(t,m_{1,\alpha},m_2)}{\sqrt{M(t,m_1,m_2)M(t,m_1^{'},m_2)}} - \frac{1}{2} \left[ \sqrt{\frac{M(t,m_1,m_2)}{M(t,m_1^{'},m_2)}} + \sqrt{\frac{M(t,m_1^{'},m_2)}{M(t,m_1,m_2)}} \right],  
    \end{aligned}
\end{equation}  

and  
\begin{equation}  
    \begin{aligned}
        &\cos(\alpha^{'}+\beta) \\
        &= \frac{M(t,m_{1,\alpha^{'}},m_2)}{\sqrt{M(t,m_1,m_2)M(t,m_1^{'},m_2)}} - \frac{1}{2} \left[ \sqrt{\frac{M(t,m_1,m_2)}{M(t,m_1^{'},m_2)}} + \sqrt{\frac{M(t,m_1^{'},m_2)}{M(t,m_1,m_2)}} \right] . 
    \end{aligned}
\end{equation}  
From these two equations, one can uniquely determine the value of $\beta \in [0,2\pi)$.

Since this relative phase is more challenging to measure in a real experiment, as it requires preparing the initial state of subsystem B as a superposition of any two basis states that form a complete basis for subsystem B, we can instead consider its lower and upper bounds and obtain a quantity, which may be easier to measure experimentally.

\subsection{The lower bound}
\label{lower_bound_appendix}
First, we consider the lower bound of the $n$-th R\'enyi entropy. From the Eq.~\eqref{n_Renyi_def}, we have
\begin{equation}
    \text{Tr}_A (\hat{\rho}_A^n) \leq   \sum_{m_1,m_2,\dots ,m_n=1}^{D_B}\prod_{j=1}^n \left|\langle \psi,\phi_j,m_{[j+1]}|   \hat{U}^{\dagger}_{j} \hat{U}_{[j+1]} |\psi,m_j,\phi_{[j+1]} \rangle \right|.
\end{equation}
Using the definition  of the basis-resolved ETP
\begin{equation}  
    M(t,m_j,m_{[j+1]}) \equiv \left| \langle \psi, \phi_j, m_{[j+1]} | \hat{U}_{A,B_j}^\dagger(t) \hat{U}_{A,B_{[j+1]}}(t) | \psi, m_j, \phi_{[j+1]} \rangle \right|^2,  
\end{equation} 
we further have
\begin{equation}
    \text{Tr}_A \left[ \hat{\rho}_A^n(t) \right] \leq   \sum_{m_1,m_2,\dots ,m_n=1}^{D_B}\prod_{j=1}^n \sqrt{M(t,m_j,m_{[j+1]})} .
\end{equation}
Thus, the $n$-th R\'enyi entropy is lower bounded by
\begin{equation}
    S_A^{(n)} \geq \frac{1}{1-n} \log\left[ \sum_{m_1,m_2,\dots ,m_n=1}^{D_B}\prod_{j=1}^n \sqrt{M(t,m_j,m_{[j+1]})}  \right].
\label{lower_bound}
\end{equation}

\subsection{The upper bound}
\label{upper_bound_appendix}
For the upper bound of the $n$-th R\'enyi entropy, one important thing to notice is that we should view 
\begin{equation}
\text{Tr}(\hat{\rho})=\int d\lambda p(\lambda)=1.
\end{equation}
Thus,
\begin{equation}
\text{Tr}(\hat{\rho}^2)=\mathbb{E}\left[\hat{\rho} \right]=\int d\lambda p(\lambda)\times p(\lambda).
\end{equation}
Here, $\mathbb{E}[\hat{O}]=\text{Tr}[\hat{\rho} \hat{O}]$. Similarly,
\begin{equation}
\text{Tr}(\hat{\rho}^n)=\mathbb{E}\left[\hat{\rho}^{n-1} \right]=\int d\lambda p(\lambda)\times p(\lambda)^{n-1}.
\end{equation}
Additionally, by applying Jensen's inequality, we obtain
\begin{equation}
G[\mathbb{E}(x)] \leq \mathbb{E}[G(x)]
\end{equation}
when choosing $G(x) = x^{n-1}$ for $0<x<1$ and $n\geq2$. Selecting $x=\lambda$, where $\lambda$ is the eigenvalue of the density matrix $\hat{\rho}_A$, gives us
\begin{equation}
[\mathbb{E}(\hat{\rho}_A)]^{n-1}\leq \mathbb{E}[\hat{\rho}_A^{n-1}],
\end{equation}
which is 
\begin{equation}
\left[\text{Tr}_A(\hat{\rho}_A^2)\right]^{n-1}\leq \text{Tr}_A\left[ \hat{\rho}_A^n \right].
\end{equation}
Thus, we have 
\begin{equation}
S_A^{(n)} \leq \frac{n-1}{1-n}\log\text{Tr}_A\left[(\hat{\rho}_A^2) \right]=S^{(2)}_A.
\label{upper_bound}
\end{equation}
Thus, the $n$-th R\'enyi entropy is upper bounded by the second R\'enyi entropy. This upper bound is not a new result that we derived for the first time; it can be inferred from the monotonicity in $n$ of the $n$-th R\'enyi entropy \cite{Beck1993-BECTOC}.

For arbitrary order $ n \geq 2 $, the second R\'enyi entropy can be computed without knowing the basis-resolved ETP distribution, providing an upper bound via Eq. \eqref{upper_bound}. If the distribution of basis-resolved ETP is measurable, the lower bound of the $ n $-th R\'enyi entropy follows from Eq. \eqref{lower_bound}.

\section{More on the diagrammatic technique for proving the OTOC-ETP relation}
\label{OTOC-LE_relation_appendix}
In this Appendix, we introduce the diagrammatic technique used to prove the OTOC-ETP relation in Section \ref{OTOC-LE_relation_main} of the main text (a similar diagrammatic proof technique can be found in \cite{fanOutofTimeOrderCorrelationManyBody2017, Mele_2024}).

\begin{figure}[ht] 
		\centering \includegraphics[width=0.95\textwidth]{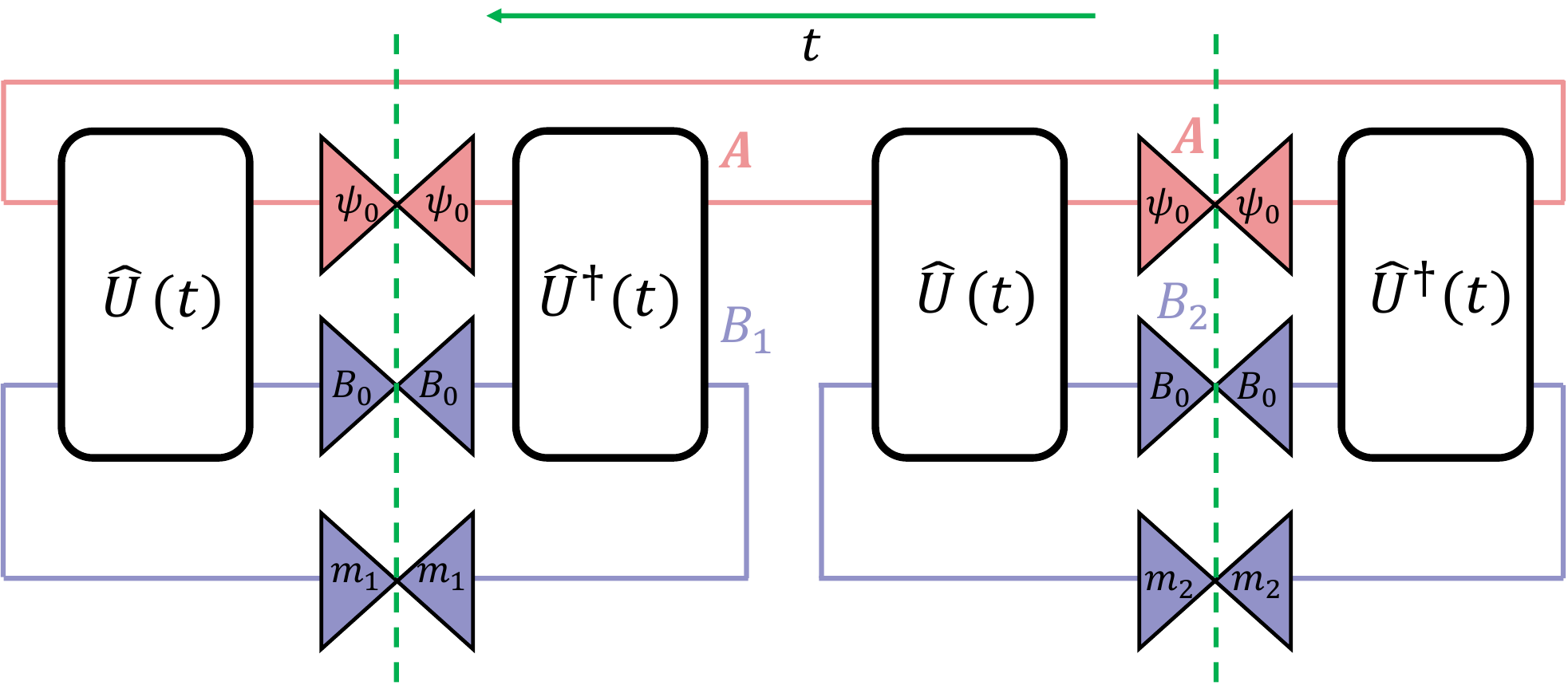} 
		\caption{The diagram representation of average OTOC defined in Eq.~\eqref{average_OTOC_A} with $\hat{W}=\hat{\rho}(0)=|\psi_0\rangle_{AA}\langle \psi_0|\otimes |B_0\rangle_{BB}\langle B_0|$. The green dashed lines are added to help clarify how the measurement protocol can be interpreted as basis-resolved ETP, and the green arrow indicates the time direction.}
		\label{RE_LE_fig}
\end{figure}

The diagram in Fig.~\ref{purity_fig} in the main text already illustrates how averaged OTOC (left-hand side of the Eq.~\eqref{OTOC-LE_equation}) can be measured as the sum of basis-resolved ETP. To further assist the Reader, we provide a guided figure, Fig.~\ref{RE_LE_fig}, to make this interpretation clearer. Compared to Fig.~\ref{purity_fig}, this figure includes green dashed lines to clarify how the measurement protocol corresponds to basis-resolved ETP, while the green arrow indicates the time direction.

In this figure, one can see that the purity of subsystem $A$ can be measured by first preparing the initial state as $ |\psi_0\rangle $ for subsystem $A$ and $ |B_0\rangle $ for subsystem $ B_2 $. Then, $A$ and $B_2$ evolve unitarily together for a time $ t $. Next, we introduce another subsystem, $ B_1 $, initialized in the state $ |m_1\rangle $. After that, $A$ and $B_1$ evolve backward together for the same time duration $ t $. Finally, we perform a projected measurement of the final state on $ |\psi_0\rangle $, $ |B_0\rangle $, and $ |m_2\rangle $ for subsystems A, $ B_1 $, and $ B_2 $, respectively.

\section{Measurement cost of the basis-resolved ETP protocol for 2-R\'enyi entropy}
\label{Measurement_cost_appendix}

This appendix estimates the measurement resources required by the basis-resolved ETP protocol introduced in
Sec.~\ref{measure_Renyi_entropy_subsection}. We distinguish three layers of cost:
(i) the total number of experimental rounds (shots),
(ii) the total number of local $\sigma_z$ readouts, and
(iii) the choice of $N_{\text{cycle}}$ needed to achieve a desired statistical accuracy in the second R\'enyi entropy.

\subsection*{Shot count}
In step~5 of the protocol in Sec.~\ref{sec:general_proposal}, for each basis label
$m_1\in\{1,\dots,D_B\}$ we repeat the single-round experiment $N_{\text{cycle}}$ times. Therefore,
the total number of experimental rounds (shots) is
\begin{equation}
N_{\text{shots}} \;=\; D_B\,N_{\text{cycle}}.
\label{eq:Nshots_appendix}
\end{equation}

\subsection*{Local readouts}
Each shot always measures subsystem $B$ in the $\sigma_z$ basis (step~3).
Subsystem $A$ is measured in the $\sigma_z$ basis (step~4) \emph{only if} the $B_1$ measurement passes.
Let $n_B$ and $n_A$ be the number of qubits in $B_1$ and $A$, respectively, and define
\begin{equation}
p_B \;\equiv\; \Pr\!\left(\text{the step~3 check yields }|+1\rangle_{B_1}\right),
\end{equation}
where $p_B$ is understood as the pass probability averaged over the full loop over $m_1$.
Then the expected total number of single-qubit $\sigma_z$ readouts is
\begin{equation}
N_{\text{loc}}
= \underbrace{n_B\,N_{\text{shots}}}_{\text{$B_1$ measured every shot}}
+ \underbrace{n_A\,p_B\,N_{\text{shots}}}_{\text{$A$ measured only if step~3 passes}}.
\label{eq:Nloc_general_appendix}
\end{equation}
In particular, this implies the bounds
\begin{equation}
n_B\,N_{\text{shots}}
\;\le\;
N_{\text{loc}}
\;\le\;
(n_A+n_B)\,N_{\text{shots}}
=
(n_A+n_B)\,D_B\,N_{\text{cycle}},
\label{eq:Nloc_bounds_appendix}
\end{equation}
where the upper bound corresponds to $p_B\simeq 1$.

\subsection*{Purity estimator and rare-event statistics}
\label{app:purity_sample_complexity}
We analyze how large $N_{\text{cycle}}$ must be to estimate the purity 
\begin{equation}
x \;\equiv\; \Tr\!\left[\hat{\rho}_A^2(t)\right]
\end{equation}
using the simplified protocol that only counts the number of ``fail'' events $N_{\text{not}}$.
The estimator obtained from the protocol is
\begin{equation}
\hat x
=
D_B-\frac{N_{\text{not}}}{N_{\text{cycle}}}.
\label{eq:xhat_from_Nnot_appendix}
\end{equation}
It is convenient to introduce the complementary ``success'' count as
\begin{equation}
N_{\text{succ}} \;\equiv\; D_B\,N_{\text{cycle}}-N_{\text{not}},
\end{equation}
for which
\begin{equation}
\hat x = \frac{N_{\text{succ}}}{N_{\text{cycle}}}.
\label{eq:xhat_from_Nsucc_appendix}
\end{equation}

For a fixed $m_1$, each shot results in either a success (no increment of $N_{\text{not}}$)
or a failure (an increment of $N_{\text{not}}$). Let $q_{m_1}$ be the success probability conditioned on $m_1$.
Since we perform $N_{\text{cycle}}$ independent shots for each $m_1$ and sum over $m_1$, we have
\begin{equation}
\mathbb{E}[N_{\text{succ}}]
=
\sum_{m_1=1}^{D_B} N_{\text{cycle}}\,q_{m_1}
=
N_{\text{cycle}}\sum_{m_1=1}^{D_B} q_{m_1}.
\end{equation}
By construction the estimator is unbiased, $\mathbb{E}[\hat x]=x$, and therefore
\begin{equation}
\sum_{m_1=1}^{D_B} q_{m_1} = x,
\qquad\Rightarrow\qquad
\mathbb{E}[N_{\text{succ}}]=N_{\text{cycle}}\,x.
\label{eq:mean_Nsucc_appendix}
\end{equation}

When the purity is small, $x\ll 1$, successes are rare and $N_{\text{succ}}$ becomes a rare-event counter.
In this regime it is natural to approximate $N_{\text{succ}}$ by a Poisson random variable with mean
\begin{equation}
\lambda \;\equiv\; N_{\text{cycle}}\,x,
\end{equation}
so that
\begin{equation}
\Var(N_{\text{succ}})\approx \lambda = N_{\text{cycle}}x,
\qquad
\sigma(N_{\text{succ}})\approx \sqrt{N_{\text{cycle}}x}.
\end{equation}
Using Eq.~\eqref{eq:xhat_from_Nsucc_appendix}, the statistical uncertainty of the purity estimator is then
\begin{equation}
\Var(\hat x)
=
\frac{\Var(N_{\text{succ}})}{N_{\text{cycle}}^2}
\approx
\frac{x}{N_{\text{cycle}}},
\qquad
\sigma(\hat x)\approx \sqrt{\frac{x}{N_{\text{cycle}}}}.
\label{eq:sigma_xhat_appendix}
\end{equation}

\paragraph{Choosing $N_{\text{cycle}}$.}
Equation~\eqref{eq:sigma_xhat_appendix} immediately yields two practical criteria:
\begin{enumerate}
\item \textbf{Detection threshold.}
To observe a nonzero signal with order-one probability one needs at least an order-one expected number of successes,
\begin{equation}
\mathbb{E}[N_{\text{succ}}]=N_{\text{cycle}}x \gtrsim 1
\qquad\Rightarrow\qquad
N_{\text{cycle}} \gtrsim \frac{1}{x}.
\label{eq:Ncycle_detect_appendix}
\end{equation}
If $N_{\text{cycle}}x\ll 1$, typically $N_{\text{succ}}=0$ and hence $\hat x=0$.

\item \textbf{Target relative accuracy.}
Requiring a relative error $\sigma(\hat x)/x\le \eta$ gives
\begin{equation}
\frac{\sigma(\hat x)}{x}
\approx
\frac{1}{\sqrt{N_{\text{cycle}}x}}
\le \eta
\qquad\Rightarrow\qquad
N_{\text{cycle}} \gtrsim \frac{1}{\eta^2\,x}.
\label{eq:Ncycle_relative_appendix}
\end{equation}
\end{enumerate}

\subsection*{Implications for highly entangled states and scaling of local readouts}
For a highly entangled state of $A$ with $n_A$ qubits, the purity is typically exponentially small, $x \sim 2^{-n_A}$. Equations~\eqref{eq:Ncycle_detect_appendix}--\eqref{eq:Ncycle_relative_appendix} imply that $N_{\text{cycle}}$
must scale exponentially with $n_A$:
\begin{equation}
N_{\text{cycle}} \gtrsim x^{-1}\sim 2^{n_A}
\quad \text{(detection)},\qquad
N_{\text{cycle}} \gtrsim \frac{2^{n_A}}{\eta^2}
\quad \text{(relative error }\eta\text{)}.
\label{eq:Ncycle_exp_scaling_appendix}
\end{equation}

Combining this with the shot count \eqref{eq:Nshots_appendix} shows that the total number of shots scales as
\begin{equation}
N_{\text{shots}}
=
D_B\,N_{\text{cycle}}
\gtrsim
\frac{D_B}{\eta^2\,x}.
\label{eq:Nshots_scaling_appendix}
\end{equation}
If $B$ consists of $n_B$ qubits and we sum over a full computational basis, then $D_B=2^{n_B}$ and
\begin{equation}
N_{\text{shots}} \sim \frac{2^{n_A+n_B}}{\eta^2}
= \frac{2^{N}}{\eta^2},
\qquad N\equiv n_A+n_B,
\label{eq:Nshots_high_ent_appendix}
\end{equation}
up to additional logarithmic factors if one fixes a confidence level.

Finally, the corresponding local-readout cost follows from Eq.~\eqref{eq:Nloc_general_appendix}:
\begin{equation}
N_{\text{loc}}
=
\bigl(n_B+n_A p_B\bigr)\,N_{\text{shots}}
\sim
\bigl(n_B+n_A p_B\bigr)\,\frac{2^{N}}{\eta^2},
\label{eq:Nloc_high_ent_appendix}
\end{equation}
with the worst-case bound (using $0\le p_B\le 1$)
\begin{equation}
n_B\,\frac{2^{N}}{\eta^2}
\;\lesssim\;
N_{\text{loc}}
\;\lesssim\;
(n_A+n_B)\,\frac{2^{N}}{\eta^2}
=
N\,\frac{2^{N}}{\eta^2}.
\label{eq:Nloc_high_ent_bounds_appendix}
\end{equation}
Thus, while the protocol is measurement-setting efficient (only $\sigma_z$ readout is required),
estimating exponentially small purities necessarily incurs an exponential cost in the total system size $N$.
The conditional readout in step~4 can reduce the prefactor by avoiding measurements of $A$ on shots that fail the
$B_1$ check, but it cannot remove the exponential scaling originating from $x\sim 2^{-n_A}$.

\section{Imperfect time-reversal error}
\label{appendix:time_reversal}
In our protocol of measuring quantum purity (second R\'enyi entropy), one demanding operation is the backward evolution. In practice, the implemented ``backward'' unitary is more accurately written as
\begin{equation}
\widetilde{U}^{\dagger}(t)=e^{+i(\hat{H}+\delta \hat{H})t}=\hat{E}(t)\,\hat{U}^{\dagger}(t),\qquad
\hat{E}(t)=\hat{\mathcal{T}}\exp\!\Big(i\int_0^t ds\,\delta \hat{H}_I(s)\Big),
\end{equation}
where $\delta \hat{H}$ captures residual couplings, calibration drift, and imperfect sign flips, and $\delta \hat{H}_I(s)=e^{+iHs}\delta \hat{H} e^{-iHs}$. We use $\widetilde U^\dagger$ to denote the imperfect time reversal. The corresponding return fidelity
\begin{equation}
\mathcal{L}(t)=\big|\langle\psi_0|\hat{E}(t)|\psi_0\rangle\big|^2
\end{equation}
directly quantifies the quality of the reversal. At short times one obtains the universal quadratic sensitivity
\begin{equation}
1-\mathcal{L}(t)=t^2\,\mathrm{Var}_{\psi_0}(\delta \hat{H})+O(t^3),
\label{eq:echo_short_time}
\end{equation}
while at longer times $\mathcal{L}(t)$ typically crosses over to exponential-type decays whose rate depends on both the dynamics and the perturbation strength \cite{Gorin2006}.

For feasibility, the crucial point is that our R\'enyi estimator is the expectation value of a bounded observable. Therefore, the deviation induced by imperfect reversal is controlled by the trace distance,
\begin{equation}
|\widetilde m(t)-m(t)|\le \tfrac12\|\widetilde\rho(t)-\hat{\rho}(t)\|_1.
\end{equation}
In the experimentally common regime where coherent unitary mismatch dominates (e.g., initial pure product states), $\|\widetilde\rho-\rho\|_1$ can be bounded by the echo infidelity via the Fuchs--van de Graaf inequality, giving a directly measurable error bar,
\begin{equation}
|\widetilde m(t)-m(t)| \;\lesssim\; \sqrt{1-\mathcal{L}(t)}.
\label{eq:signal_bound_by_echo}
\end{equation}
Since $S^{(2)}_A=-\ln\!\big(\mathrm{Tr}\,\hat{\rho}_A^2\big)$, an absolute purity error $\Delta P$ (with $P=\mathrm{Tr}\,\hat{\rho}_A^2$) propagates as
\begin{equation}
|\Delta S^{(2)}_A|\approx \frac{|\Delta P|}{P}.
\end{equation}
Hence the long-time, low-purity regime is the most demanding: one must keep the absolute purity error small compared to $P$. This limitation is generic for echo-based probes of scrambling/OTOCs, not specific to our protocol.

To make these bounds operational, one can first measure the \emph{benchmark} $\mathcal{L}(t)$ on the same hardware by running the identical forward/backward sequence with the intermediate operations removed, as routinely done in experimental echo/OTOC demonstrations. This yields (i) a platform-specific ``usable time window'' where $\mathcal{L}(t)$ remains high and Eq.~\eqref{eq:signal_bound_by_echo} ensures controlled systematics, and (ii) a direct characterization of contrast loss that can be incorporated into conservative error bars. Such echo benchmarking and error attribution are standard in trapped-ion, NMR, and superconducting-circuit time-reversal experiments \cite{Gaerttner2017,Sanchez2020,Braumuller2022,Zhao2022,Swingle2016}.


\end{document}